\pdfoutput=1

\documentclass[11pt,twoside,a4paper,cmspaper,final,collab]{cms-tdr}
\def\svnVersion{292307}\def\cmsCernNo{2013-037}\def\cmsCernDate{\today}\def\cmsMessage{Published in the European Physical Journal C as \href{http://dx.doi.org/10.1140/epjc/s10052-015-3454-1}{\doi{10.1140/epjc/s10052-015-3454-1.}}}

\begin{document}\cmsNoteHeader{HIG-14-010}

\hyphenation{had-ron-i-za-tion}
\hyphenation{cal-or-i-me-ter}
\hyphenation{de-vices}

\RCS$Revision: 286889 $
\RCS$HeadURL: svn+ssh://svn.cern.ch/reps/tdr2/papers/HIG-14-010/trunk/HIG-14-010.tex $
\RCS$Id: HIG-14-010.tex 286889 2015-04-29 20:51:10Z alverson $
\newlength\cmsFigWidth
\ifthenelse{\boolean{cms@external}}{\setlength\cmsFigWidth{0.85\columnwidth}}{\setlength\cmsFigWidth{0.4\textwidth}}
\ifthenelse{\boolean{cms@external}}{\providecommand{\cmsLeft}{top}}{\providecommand{\cmsLeft}{left}}
\ifthenelse{\boolean{cms@external}}{\providecommand{\cmsRight}{bottom}}{\providecommand{\cmsRight}{right}}
\ifthenelse{\boolean{cms@external}}{\providecommand{\cmsTopLeft}{top}}{\providecommand{\cmsTopLeft}{top left}}
\ifthenelse{\boolean{cms@external}}{\providecommand{\cmsMiddle}{middle}}{\providecommand{\cmsMiddle}{top right}}

\ifthenelse{\boolean{cms@external}}{\providecommand{\CL}{CL\xspace}}{\providecommand{\CL}{CL\xspace}}
\newcommand{\CLs}{\ensuremath{\mathrm{CL}_\mathrm{s}}\xspace}
\ifthenelse{\boolean{cms@external}}{\providecommand{\NA}{\ensuremath{\cdots}}}{\providecommand{\NA}{\text{---}}}

\providecommand{\qqbar}{\ensuremath{\PQq\PAQq}\xspace}
\newcommand{\jj}{\ensuremath{\cmsSymbolFace{jj}}}
\newcommand{\ttJJ}{\ensuremath{\ttbar\text{+jet}}}
\newcommand{\ttJJs}{\ensuremath{\ttbar\text{+jets}}}
\newcommand{\WJJ}{\ensuremath{\PW\text{+jets}}}
\newcommand{\ZJJ}{\ensuremath{\Z\text{+jets}}}
\newcommand{\ttH}{\ensuremath{\ttbar\PH}}
\newcommand{\ttZ}{\ensuremath{\ttbar\Z}}
\newcommand{\ttW}{\ensuremath{\ttbar\PW}}
\newcommand{\ttWZ}{\ensuremath{\ttbar\mathrm{+}\PW/\Z}}
\newcommand{\ttHBB}{\ensuremath{\ttbar\PH(\bbbar)}}
\newcommand{\HBB}{\ensuremath{\PH\to\bbbar}}
\newcommand{\HZZ}{\ensuremath{\PH\to\Z\Z^*}}
\newcommand{\HWW}{\ensuremath{\PH\to\PW\PW^*}}
\newcommand{\ggH}{\ensuremath{\Pg\Pg\to\PH}}
\newcommand{\HTT}{\ensuremath{\PH\to\TT}}
\newcommand{\ttBB}{\ensuremath{\ttbar\bbbar}}
\newcommand{\ttUU}{\ensuremath{\ttbar\jj}}
\newcommand{\ttpB}{\ensuremath{\ttbar\mathrm{+}\PQb}}
\newcommand{\ttpBB}{\ensuremath{\ttbar\mathrm{+}\bbbar}}
\newcommand{\ttpBBbar}{\ensuremath{\ttbar\mathrm{+}\bbbar}}
\newcommand{\ttpCC}{\ensuremath{\ttbar\mathrm{+}\PQc\PQc}}
\newcommand{\ttpCCbar}{\ensuremath{\ttbar\mathrm{+}\ccbar}}
\newcommand{\ttpJJ}{\ensuremath{\ttbar\mathrm{+}\jj}}
\newcommand{\ttpLF}{\ensuremath{\ttbar\mathrm{+lf}}}
\newcommand{\ttpHF}{\ensuremath{\ttbar\mathrm{+hf}}}
\newcommand{\WQQ}{\ensuremath{\PW\to\qqbar}}
\newcommand{\yT}{\ensuremath{y_{\mathrm{T}}}}
\newcommand{\mH}{\ensuremath{m_\PH}}
\newcommand{\Njet}{\ensuremath{N_\text{jet}}}
\providecommand{\PWpm}{\ensuremath{\PW^\pm}\xspace}

\cmsNoteHeader{HIG-14-010}
\title{Search for a standard model Higgs boson produced in association with a top-quark pair and decaying to bottom quarks using a matrix element method}
\titlerunning{Search for Higgs in association with \PQt to \PQb quarks using matrix elements}

\date{\today}

\abstract{
A search for a standard model Higgs boson produced in association with a top-quark pair
and decaying to bottom quarks is presented.
Events with hadronic jets and one or two oppositely charged leptons are selected from a data sample corresponding
to an integrated luminosity of 19.5\fbinv collected by the CMS experiment at the LHC in $\Pp\Pp$
collisions at a centre-of-mass energy of 8\TeV.
In order to separate the signal from the larger \ttbar+jets background,
this analysis uses a matrix element method that assigns a probability density value to each reconstructed event under
signal or background hypotheses.
The ratio between the two values
is used in a maximum likelihood fit to extract the signal yield. The results are presented in terms of the measured signal strength modifier, $\mu$, relative to the standard model prediction for a Higgs boson mass of 125\GeV.
The observed (expected) exclusion limit at a 95\% confidence level is $\mu<4.2$ (3.3), corresponding to a best fit value $\hat{\mu}=1.2^{+1.6}_{-1.5}$.
}

\hypersetup{%
pdfauthor={CMS Collaboration},%
pdftitle={Search for a standard model Higgs boson produced in association with a top-quark pair and decaying to bottom quarks using a matrix element method},%
pdfsubject={CMS},%
pdfkeywords={CMS, physics, top, matrix element}}

\maketitle

\section{Introduction} \label{sec:introduction}

Following the discovery of a new boson with mass around 125\GeV
by the ATLAS and CMS Collaborations~\cite{Higgs2,Higgs1,Chatrchyan:2013lba} at the CERN LHC,
the measurement of its properties has become an important task in particle physics. The precise determination
of its quantum numbers and couplings to gauge bosons and fermions will answer the question whether
the newly discovered particle is the Higgs boson ($\PH$) predicted by the standard model (SM) of particle physics, \ie the quantum of
the field responsible for the spontaneous breaking of the electroweak symmetry~\cite{Englert:1964et,Higgs:1964ia,Higgs:1964pj,Guralnik:1964eu,Higgs:1966ev,Kibble:1967sv}.
Conversely, any deviation from SM predictions will represent evidence of physics beyond our present knowledge, thus
opening new horizons in high-energy physics.
While the measurements performed with the data collected so far indicate overall consistency with the SM
expectations~\cite{Chatrchyan:2013lba, Chatrchyan:2012jja, Aad:2013wqa, Aad:2013xqa, Aaltonen:2013kxa},
it is necessary to continue improving on the measurement of all possible observables.

\begin{figure*}[htbp]
\centering{
  \includegraphics[width=0.32\linewidth]{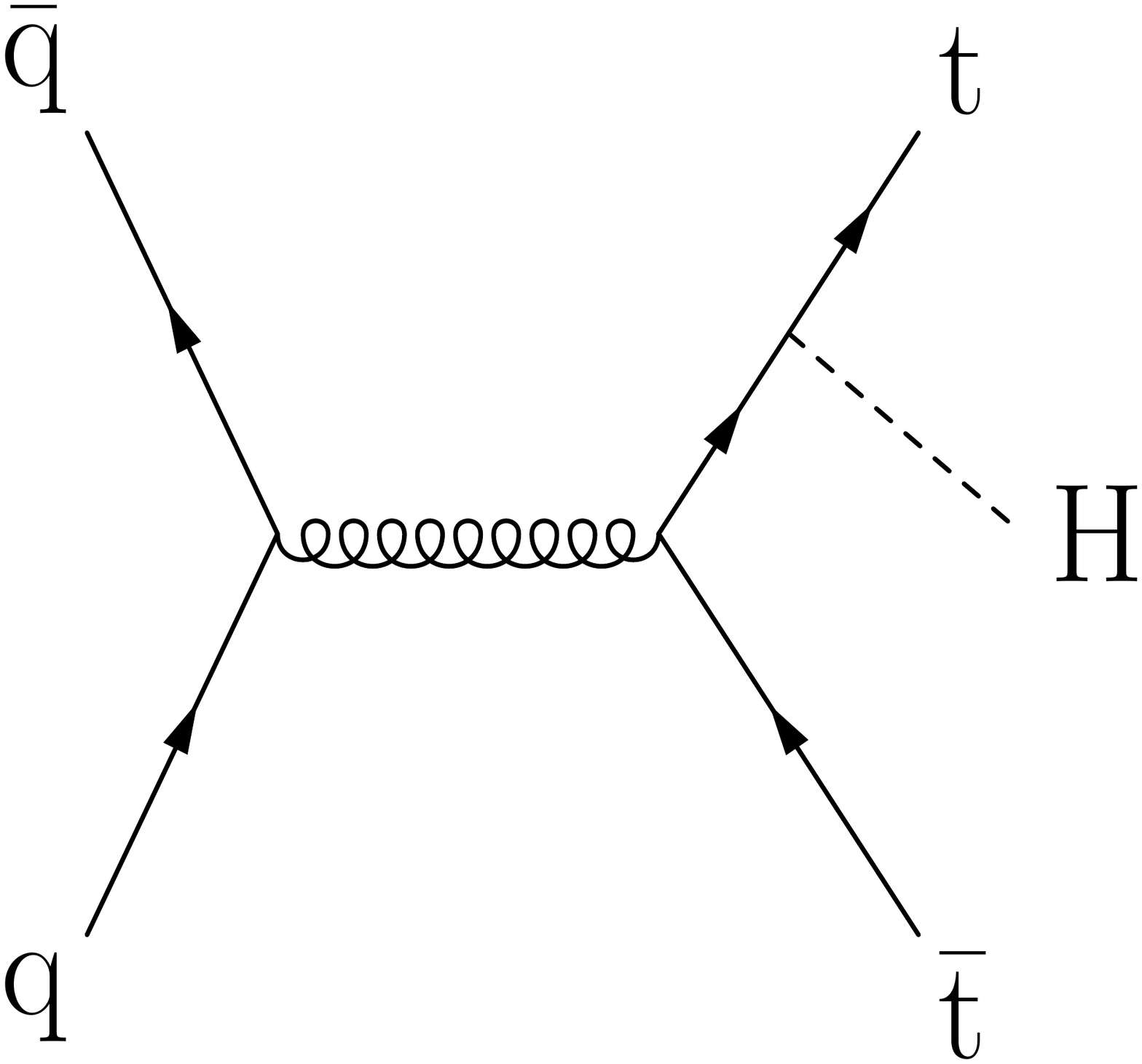}
  \includegraphics[width=0.32\linewidth]{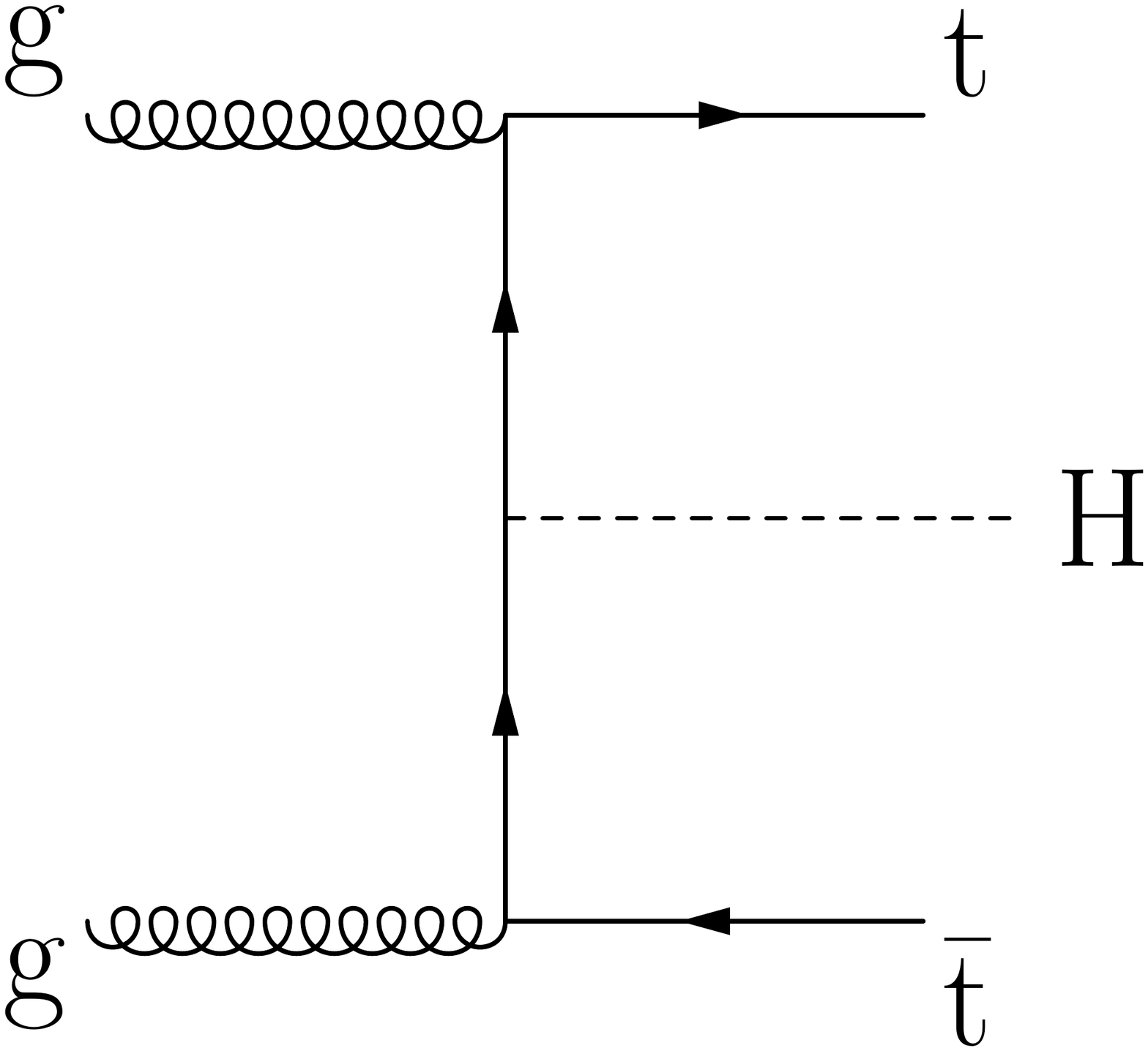}
  \includegraphics[width=0.32\linewidth]{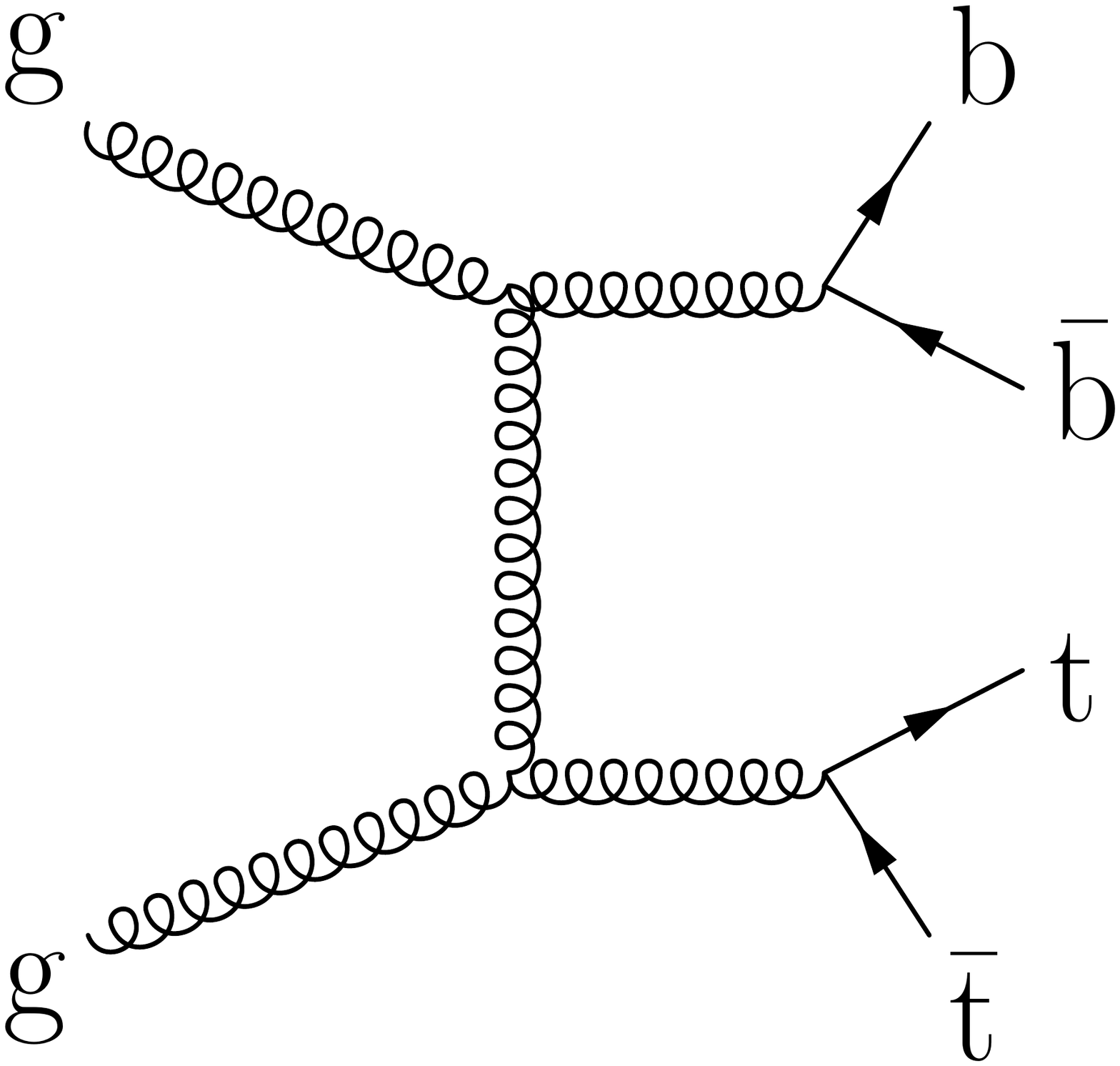}
\caption{Tree-level Feynman diagrams contributing to the partonic processes: (left) $\qqbar\to \ttH$,
(centre) $\Pg\Pg\to \ttH$, and (right) $\Pg\Pg\to \ttpBBbar$.}
\label{fig:fd}
}
\end{figure*}

{\tolerance=400
In the SM, the Higgs boson couples to fermions via Yukawa interactions with strength proportional to the fermion mass.
Direct measurements of decays into bottom quarks and $\tau$ leptons
have provided the first evidence that the 125\GeV Higgs boson couples to down-type fermions with SM-like strength~\cite{Chatrchyan:2014fermions}.
Evidence of a direct coupling to up-type fermions, in particular to top quarks, is still lacking.
Indirect constraints on the top-quark Yukawa coupling can be inferred from measuring either
the production or the decay of Higgs bosons through effective couplings generated by top-quark loops.
Current measurements of the Higgs boson cross section via gluon fusion and of its branching fraction to photons are consistent with the SM expectation for the top-quark Yukawa coupling~\cite{Chatrchyan:2013lba, Chatrchyan:2012jja, Aad:2013wqa, Aad:2013xqa}.
Since these effective couplings occur at the loop level, they can be affected by beyond-standard model (BSM) particles.
In order to disentangle the top-quark Yukawa coupling from a possible BSM contribution,
a direct measurement of the former is required.
This can be achieved by measuring observables that probe the top-quark Yukawa interaction with the Higgs boson already at the tree-level.
The production cross section of the Higgs boson in association with a top-quark pair ($\ttH$) provides an example of such an observable.
A sample of tree-level Feynman diagrams contributing to the partonic processes
$\qqbar,\Pg\Pg\to \ttH$ is shown in Fig.~\ref{fig:fd} (left and centre).
The inclusive next-to-leading-order
(NLO) $\ttH$ cross section is about $130\unit{fb}$ in $\Pp\Pp$ collisions at a centre-of-mass
energy $\sqrt{s}=8\TeV$ for a Higgs boson mass ($\mH$) of 125\GeV
~\cite{Raitio:1978pt, Ng:1983jm,
  Kunszt:1984ri, Beenakker:2001rj, Beenakker:2002nc, Dawson:2002tg,
  Dawson:2003zu,Garzelli:2011vp,YellowReport,YellowReport3}, which is
approximately two orders of magnitude smaller than the cross section for Higgs boson production via gluon fusion~\cite{YellowReport,YellowReport3}.
\par}

The first search for $\ttH$ events used $\Pp\Pap$ collision data at
$\sqrt{s}=1.96\TeV$ collected by the CDF experiment at the Tevatron collider~\cite{cdf_tth}.
Searches for $\ttH$ production at the LHC have previously been published
for individual decay modes of the Higgs boson~\cite{PAPERttHAtlas,PAPERttHbb}.
The first combination of $\ttH$ searches in different final states has been published by the CMS Collaboration
based on the full data set collected at $\sqrt{s}=7$ and 8\TeV~\cite{PAPERttHCMS}.
Assuming SM branching fractions, the results of that analysis
set a 95\% confidence level (\CL) upper limit on the $\ttH$ signal strength at 4.5 times the SM value,
while an upper limit of 1.7 times the SM is expected from the background-only hypothesis.
The median expected exclusion limit for $\ttH$ production in the $\HBB$ channel alone is 3.5
in the absence of a signal.

The results of a search for $\ttH$ production in the decay channel $\HBB$ are presented in this paper
based on $\Pp\Pp$ collision data at $\sqrt{s}=8\TeV$ collected with the CMS detector~\cite{CMS} and
corresponding to an integrated luminosity of 19.5\fbinv.
The analysis described here differs from that of Ref.~\cite{PAPERttHCMS} in the way events are categorized
and in its use of an analytical matrix element method (MEM)~\cite{MEM1,MEM2} for improving the separation
of signal from background.
Within the MEM technique,
each reconstructed event is assigned a probability density value
based on the theoretical differential cross section $\sigma^{-1}\rd\sigma/\rd\vec{y}$,
where $\vec{y}$ denotes the four-momenta of the reconstructed particles.
Particle-level quantities that are either unknown (\eg neutrino momenta, jet-parton associations) or
poorly measured (\eg quark energies) are marginalised by integration.
The ratio between the probability density values for signal and background provides a discriminating variable suitable
for testing the compatibility of an event with either of the two hypotheses~\cite{Neyman}.

The MEM has already been successfully used at the Tevatron collider in the context of Higgs boson searches~\cite{MEM8,MEM9}, although
for simpler final states.
A phenomenological feasibility study for a $\ttH$ measurement in the $\HBB$ decay channel at the LHC using the MEM has been pioneered in Ref.~\cite{ttHMW} based on the \textsc{MadWeight} package~\cite{MW} for automatised matrix-element calculations.
The present paper makes use of an independent implementation of the MEM, specifically optimized
for the final state of interest. This is the first time that the MEM is applied to a search for $\ttH$ events.
The final states typical of $\ttH$ events with $\HBB$, that are characterised by huge combinatorial background,
the presence of nonreconstructed particles, and small signal-to-background ratios,
provide an ideal case for the deployment of the MEM.
The analysis strategy is designed to maximise the separation between $\ttH$ and $\ttpBBbar$ background events,
in order to reduce the systematic uncertainty on the signal extraction related to the modelling of this challenging background.

This paper is organised as follows. Section~\ref{sec:CMS} describes the main features of the CMS detector.
Section~\ref{sec:data} presents the data and simulation samples, while Sections~\ref{sec:obj} and~\ref{sec:selection} discuss the reconstruction of physics objects and the event selection, respectively. Section~\ref{sec:algo} describes the signal extraction.
The treatment of systematic uncertainties and the statistical interpretation of the results are discussed in Sections~\ref{sec:systematics} and~\ref{sec:results}, respectively. Section~\ref{sec:end} summarises the results.

\section{CMS detector}\label{sec:CMS}

The central feature of the CMS apparatus is a superconducting solenoid of 6\unit{m} internal diameter, providing a magnetic field of 3.8\unit{T}.
Within the field volume are a silicon pixel and strip tracker, a lead tungstate crystal electromagnetic calorimeter (ECAL),
and a brass and scintillator hadron calorimeter (HCAL), each composed of a barrel and two endcap sections.
Muons are measured in gas-ionization detectors embedded in the steel flux-return yoke outside the solenoid.
Extensive forward calorimetry complements the coverage provided by the barrel and endcap detectors.
The first level of the CMS trigger system, composed of custom hardware processors, uses information from the calorimeters and muon detectors to select the most interesting events in a time interval of less than 4\mus. The high-level trigger processor farm further decreases the event rate from around 100\unit{kHz} to around 1\unit{kHz}, before data storage.
A more detailed description of the CMS detector, together with a definition of the coordinate system used and the relevant kinematic variables can be found in Ref.~\cite{CMS}.
\section{Data and simulated samples} \label{sec:data}

The data sample used in this search was collected with the CMS detector in 2012 from $\Pp\Pp$ collisions
at a centre-of-mass energy of 8\TeV, using single-electron, single-muon, or dielectron triggers.
The single-electron trigger requires the presence of an isolated electron with transverse momentum (\pt)
in excess of 27\GeV. The single-muon trigger requires an isolated muon candidate with \pt above 24\GeV.
The dielectron trigger requires two isolated electrons with \pt thresholds of 17 and 8\GeV.

Signal and background processes are modelled with Monte Carlo (MC) simulation programs.
The CMS detector response is simulated by using the \GEANTfour software package~\cite{GEANT}.
Simulated events are required to pass the same trigger selection and offline reconstruction algorithms
used on collision data.
Correction factors are applied to the simulated samples to account for residual differences in the selection and
reconstruction efficiencies with respect to those measured.

{\tolerance=400
The $\ttH$, $\HBB$ signal is modelled by using the \PYTHIA~6.426~\cite{PYTHIA} leading order (LO)
event generator normalised to the NLO
theoretical cross section~\cite{Raitio:1978pt, Ng:1983jm,
  Kunszt:1984ri, Beenakker:2001rj, Beenakker:2002nc, Dawson:2002tg,
  Dawson:2003zu,Garzelli:2011vp,YellowReport,YellowReport3}, and assuming the SM Higgs boson with a mass of 125\GeV.
The main background in the analysis stems from $\ttJJ$ production.
This process has been simulated with the \MADGRAPH~5.1.3~\cite{MG5} tree-level matrix element generator
matched to \PYTHIA for the parton shower description, and normalised to the
inclusive next-to-next-to-leading-order (NNLO) cross section with soft-gluon resummation at next-to-next-to-leading
logarithmic accuracy~\cite{xSec_ttbar}. The $\ttJJs$ sample has been generated in a five-flavour scheme with tree-level diagrams
for two top quarks plus up to three extra partons, including both charm and bottom quarks.
An additional correction factor to the $\ttJJs$ samples is applied to
account for the differences observed in the top-quark \pt spectrum when comparing the \MADGRAPH simulation with
data~\cite{TopPtCMSPaper}.
The interference between the $\ttH$, $\HBB$ diagrams and
the $\ttpBBbar$ background diagrams is negligible and is not considered in the MC simulation.
Minor backgrounds come from the Drell--Yan production of an electroweak boson with additional jets ($\WJJ$, $\ZJJ$),
and from the production of a top-quark pair in association with a $\PWpm,\Z$ boson ($\ttW$, $\ttZ$).
These processes have been generated by \MADGRAPH matched to the \PYTHIA parton shower description.
The Drell--Yan processes have been normalised to the NNLO inclusive cross section from \FEWZ~3.1~\cite{FEWZ},
while the NLO calculations from Refs.~\cite{xSec_ttZ} and \cite{xSec_ttW} are used to normalise
the $\ttW$ and $\ttZ$ samples, respectively.
Single top quark production is modelled with
the NLO generator \POWHEG~1.0~\cite{Nason:2004rx,Frixione:2007vw,Alioli:2010xd,Melia:2011tj,Re:2010bp,Alioli:2009je}
combined with \PYTHIA.
Electroweak diboson processes ($\PW\PW$, $\PW\Z$, and $\Z\Z$) are simulated
by using the \PYTHIA generator normalised to the NLO cross section calculated with \MCFM~6.6~\cite{MCFM}.
Processes that involve top quarks have been generated with a top-quark mass of 172.5\GeV.
Samples generated at LO use the CTEQ6L1 parton distribution function (PDF) set~\cite{Pumplin:2002vw}, while samples
generated with NLO programs use the CTEQ6.6M PDF set~\cite{PhysRevD.78.013004}.
\par}

Effects from additional $\Pp\Pp$ interactions in the same bunch crossing (pileup) are modelled by
adding simulated minimum bias events (generated with \PYTHIA) to the generated hard interactions.
The pileup multiplicity in the MC simulation is reweighted to reflect the luminosity profile
observed in $\Pp\Pp$ collision data.
\section{Event reconstruction} \label{sec:obj}

The global event reconstruction provided by the particle-flow (PF) algorithm~\cite{CMS-PAS-PFT-09-001,CMS-PAS-PFT-10-001}
seeds the reconstruction of the physics objects deployed in the analysis.
To minimise the impact of pileup, charged particles are required to originate from the primary
vertex, which is identified as the reconstructed vertex with the largest value of $\sum p_{\mathrm{T},i}^2$,
where $p_{\mathrm{T},i}$ is the transverse momentum of the $i$th charged particle associated with the vertex.
The missing transverse momentum vector $\ptvecmiss$ is defined as the negative vector sum
of the transverse momenta of all neutral particles and of the charged particles
coming from the primary vertex. Its magnitude is referred to as $\ETmiss$.

Muons are reconstructed from a combination of measurements in the silicon tracker and in the muon system~\cite{Chatrchyan:2012xi}.
Electron reconstruction requires the matching
of an energy cluster in the ECAL with a track in the silicon tracker~\cite{Baffioni:2006cd}.
Additional identification criteria are applied to muon and electron candidates to reduce instrumental backgrounds.
An isolation variable is defined starting from the scalar \pt sum of all
particles contained inside a cone around the track direction,
excluding the contribution from the lepton itself.
The amount of neutral pileup energy
is estimated as the average \pt density calculated from all neutral particles in the event multiplied by an effective area of the isolation cone,
and is subtracted from the total sum.

Jets are reconstructed by using the
anti-\kt clustering algorithm~\cite{antikt}, as implemented in the \textsc{FastJet}
package~\cite{Cacciari:fastjet1,Cacciari:fastjet2},
with a distance parameter of 0.5.
Each jet is required to have pseudorapidity ($\eta$) in the range $[-2.5,2.5]$,
to have at least two tracks associated with it,
and to have electromagnetic and hadronic energy fractions of at least 1\% of the total jet
energy. Jet momentum is determined as the vector sum of the momenta of all particles in the jet.
An offset correction is applied to take into account the extra energy clustered in jets because of pileup. Jet energy corrections are derived from the simulation, and are confirmed with in situ measurements of the energy balance of dijet and $\Z/\gamma$+jet events~\cite{cmsJEC}. Additional selection criteria are applied to each event to remove spurious jet-like features originating from isolated noise patterns in few HCAL regions.

The combined secondary vertex (CSV) \PQb-tagging
algorithm is used to identify jets originating from the hadronisation of bottom quarks~\cite{Chatrchyan:2012jua}.
This algorithm combines the information about track impact parameters and secondary
vertices within jets into a likelihood discriminant to provide separation
of \PQb-quark jets from jets that originate from lighter quarks or gluons.
The CSV algorithm assigns to each jet a continuous value that can be used as a jet flavour discriminator.
Large values of the discriminator correspond preferentially to \PQb-quark jets, so that
working points of increasing purity can be defined by requiring higher values of the CSV discriminator.
For example, the CSV medium working point (CSVM) is defined in such a way as to
provide an efficiency of about 70\% (20\%) to tag jets originating
from a bottom (charm) quark,
and of approximately 2\% for jets originating from light quarks or gluons.
Scale factors are applied to the simulation to match the distribution of the CSV discriminator measured with a tag-and-probe technique~\cite{CMS_WZ}
in data control regions.
The scale factors have been derived as a function of the jet flavour, $\pt$, and $\abs{\eta}$,
as described in Ref.~\cite{PAPERttHCMS}.

\section{Event selection}\label{sec:selection}

\begin{figure*}[htbp]
\centering{
  \includegraphics[width=0.475\linewidth]{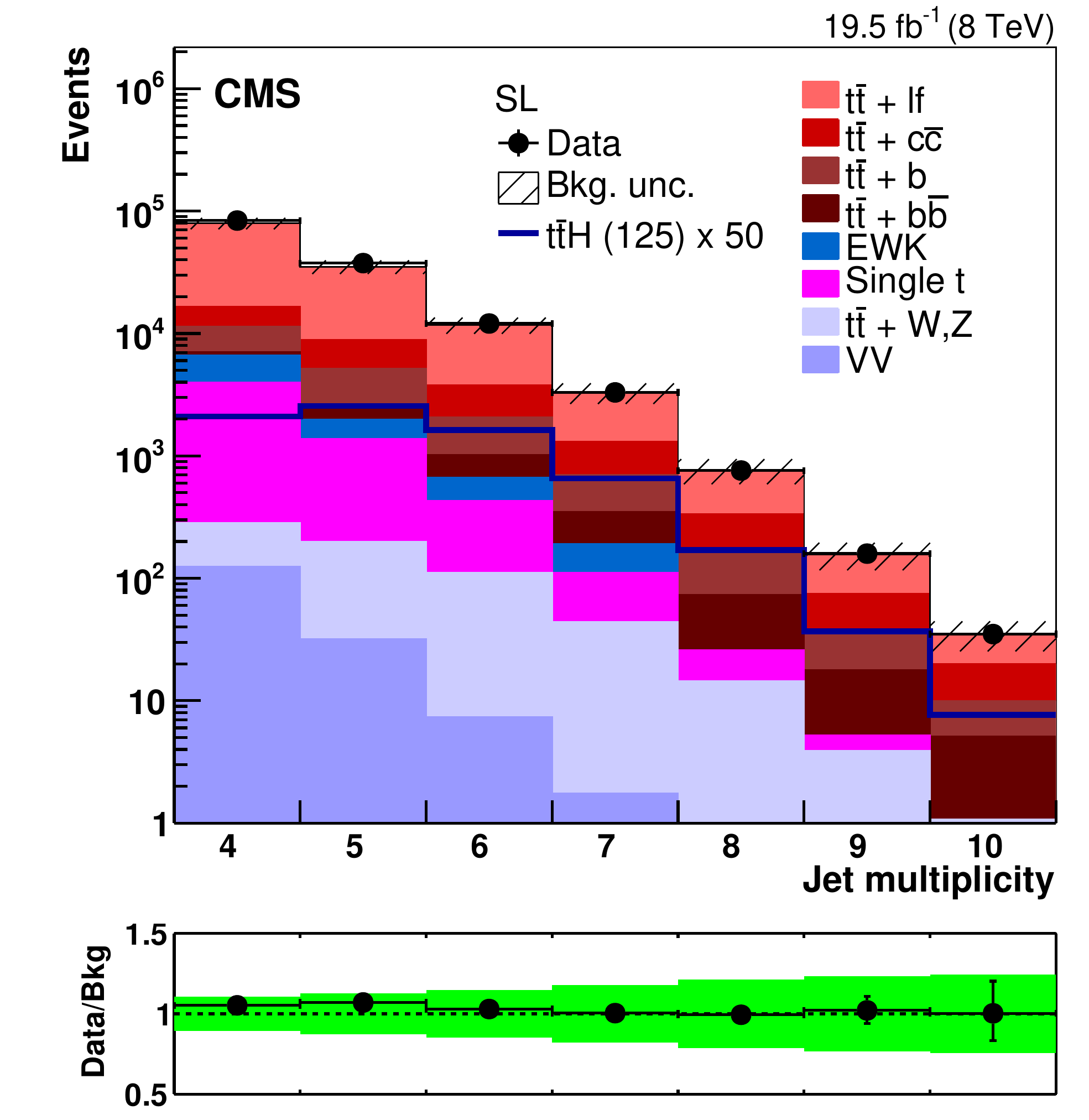}
  \includegraphics[width=0.475\linewidth]{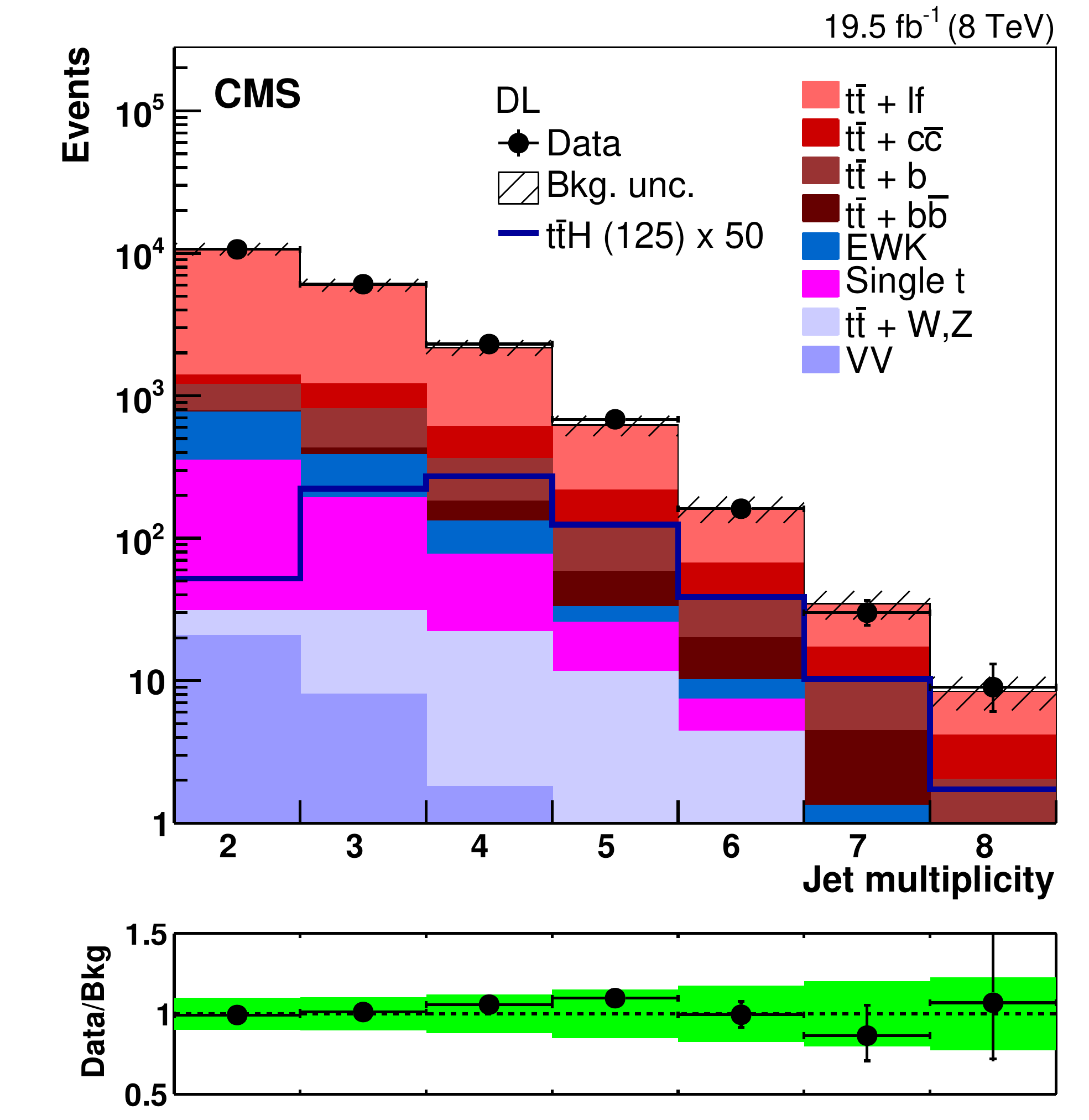}\\
  \includegraphics[width=0.475\linewidth]{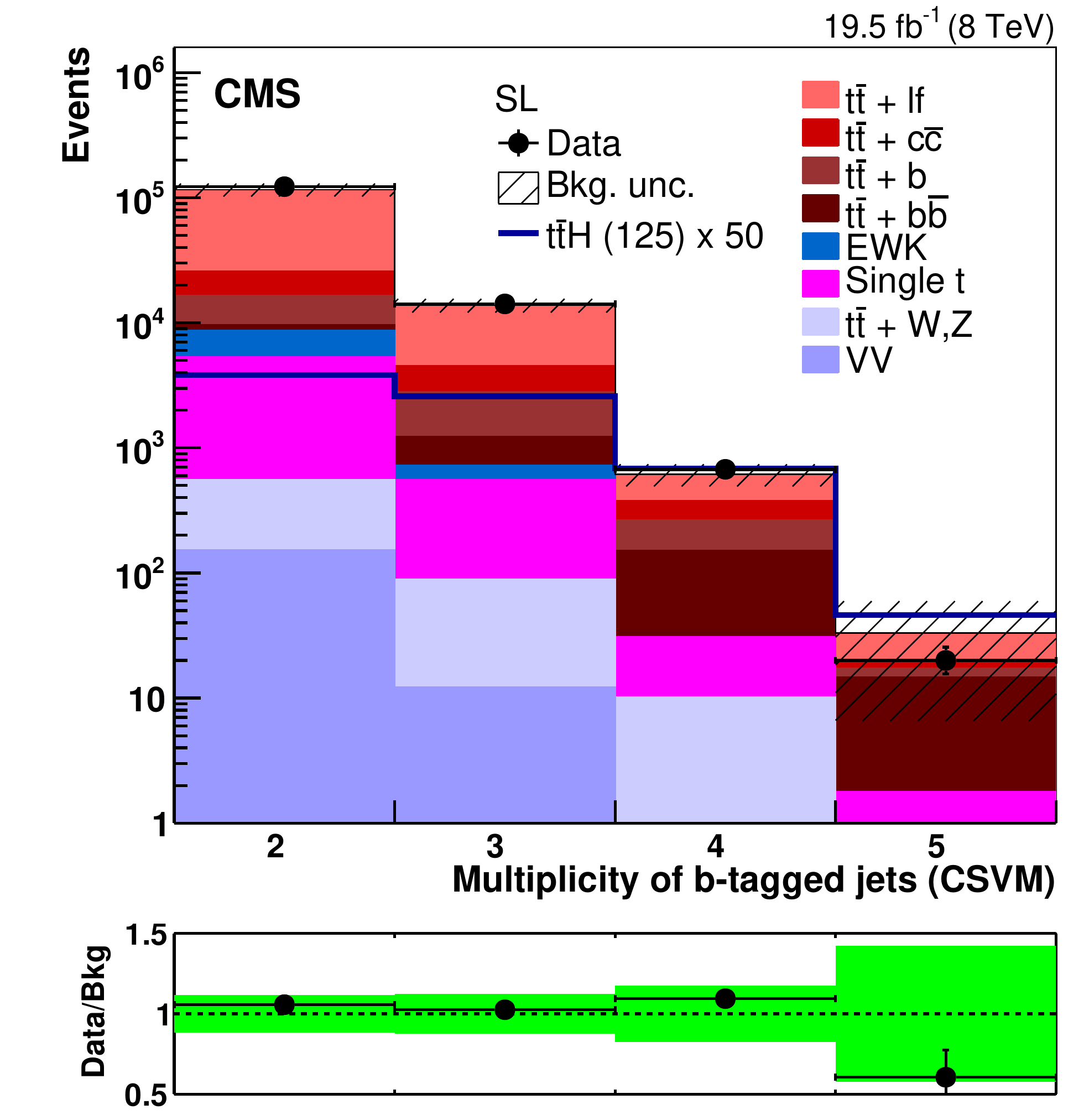}
  \includegraphics[width=0.475\linewidth]{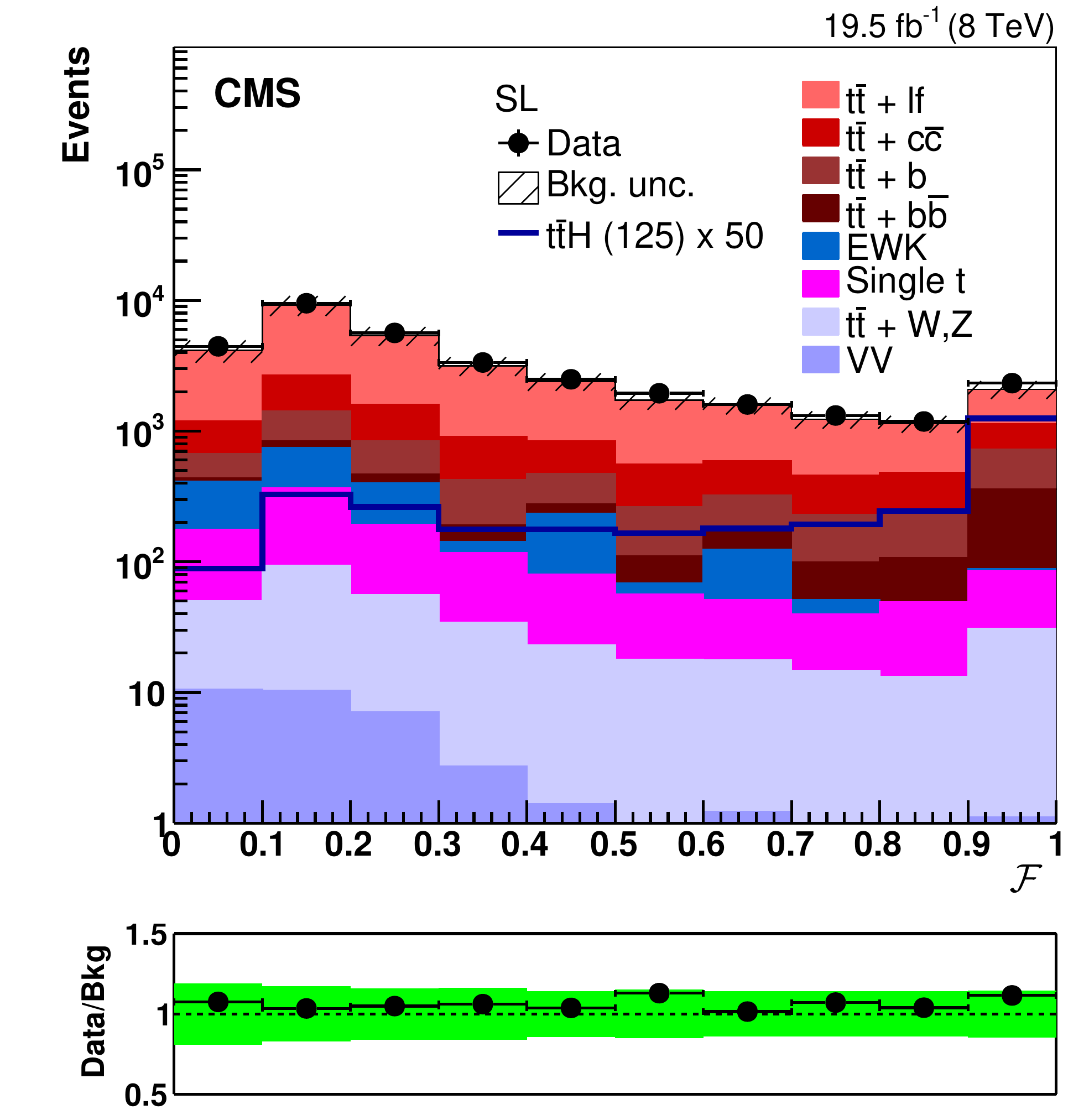}\\
\caption{Top row: distribution of the jet multiplicity in (left) single-lepton and (right) dilepton events,
after requiring that at least two jets pass the CSVM working point.
Bottom-left: distribution of the multiplicity of jets passing the CSVM working point in single-lepton events with at least four jets.
Bottom-right: distribution of the selection variable $\mathcal{F}$ defined in Eq.~\eqref{eq:btagLR} for single-lepton events with at least six jets
after requiring a loose preselection of at least one jet passing the CSVM working point.
The plots at the bottom of each panel show the ratio between the observed data and the background expectation predicted by the simulation.
The shaded and solid green bands corresponds to the total statistical plus systematic uncertainty in the background expectation described in Section~\ref{sec:systematics}.
More details on the background modelling are provided in Section~\ref{subsec:algo3}.}
\label{fig:numJets}
}
\end{figure*}

The experimental signature of $\ttH$ events with $\HBB$ is affected by a large multijet background
which can be reduced to a negligible level by only considering the semileptonic decays of the top quark.
The selection criteria are therefore optimised to accept events compatible with a $\ttH$ signal where $\HBB$ and
at least one of the top quarks decays to a bottom quark, a charged lepton, and a neutrino.
Events are divided into two exclusive channels depending on the number of charged leptons
(electrons or muons), which can be either one or two. Top quark decays in final states with tau leptons are not
directly searched for, although they can still satisfy the event selection criteria when the tau lepton
decays to an electron or muon, plus neutrinos. Channels of different lepton multiplicities are analysed separately.
The single-lepton (SL) channel requires one isolated muon with $\pt>30\GeV$ and $\abs{\eta} < 2.1$,
or one isolated electron with $\pt>30\GeV$ and $\abs{\eta} < 2.5$,
excluding the $1.44 < \abs{\eta} < 1.57$ transition region between the ECAL barrel and endcap.
Events are vetoed if additional electrons or muons with $\pt$ in excess of 20\GeV, the same $\abs{\eta}$ requirement,
and passing some looser identification and isolation criteria are found.
The dilepton (DL) channel collects events with a pair of oppositely charged leptons
satisfying the selection criteria used to veto additional leptons in the SL channel.
To reduce the contribution from Drell--Yan events in the same-flavour DL channel,
the invariant mass of the lepton pair is required to be larger than 15\GeV and at least 8\GeV away from the $\Z$ boson mass.
Figure~\ref{fig:numJets} (top) shows the jet multiplicity in the SL (left) and DL (right) channels,
while the bottom left panel of the same figure shows the multiplicity of jets
passing the CSVM working point in the SL channel.

The optimisation of the selection criteria in terms
of signal-to-background ratio requires a stringent demand on the number of jets.
At least five (four) jets with $\pt>30\GeV$ and $\abs{\eta}<2.5$ are requested in the SL (DL) channel.
A further event selection is required to reduce the $\ttJJs$ background, which at this stage exceeds the signal rate by more than three orders of magnitude.
For this purpose, the CSV discriminator values are calculated for all jets in the event and collectively denoted by $\vec{\xi}$.
For SL (DL) events with seven or more (five or more) jets, only the six (four) jets with the largest CSV discriminator value are considered.
The likelihood to observe $\vec{\xi}$ is then evaluated under the alternative hypotheses of \ttbar plus two heavy-flavour jets ($\ttpHF$) or \ttbar plus two light-flavour jets ($\ttpLF$).
For example, for SL events with six jets, and neglecting correlations among different jets in the same event,
the likelihood under the $\ttpHF$ hypothesis is estimated as:
\ifthenelse{\boolean{cms@external}}{
\begin{multline} \label{eq:btagLR_s}
f( \vec{\xi} | \ttpHF ) =
\sum_{i_1} \sum_{i_2\neq i_1} \hdots \\ \sum_{i_6\neq i_1,\hdots,i_5}
 \Bigl\{  \prod_{ k\in\{ i_1,i_2,i_3,i_4\} }\hspace*{-1.5em} f_{\mathrm{hf}}(\xi_{k})
  \prod_{ m\in \{ i_5,i_6\} } \hspace*{-0.5em}f_{\mathrm{lf}}(\xi_{m}) \Bigr\},
\end{multline}
}{
\begin{equation} \label{eq:btagLR_s}
f( \vec{\xi} | \ttpHF ) =
\sum_{i_1} \sum_{i_2\neq i_1} \hdots  \sum_{i_6\neq i_1,\hdots,i_5} \Bigl\{  \prod_{ k\in\{ i_1,i_2,i_3,i_4\} }\hspace*{-1.5em} f_{\mathrm{hf}}(\xi_{k})  \prod_{ m\in \{ i_5,i_6\} } \hspace*{-0.5em} f_{\mathrm{lf}}(\xi_{m}) \Bigr\},
\end{equation}
}
where $\xi_{i}$ is the CSV discriminator for the $i$th jet, and $f_{\mathrm{hf}\mathrm{(lf)}}$
is the probability density function (pdf) of $\xi_{i}$ when the $i$th jet originates from heavy- (light-) flavour partons.
The latter include $\PQu$, $\PQd$, $\PQs$ quarks and gluons, but not $\PQc$ quarks.
For the sake of simplicity, the likelihood in Eq.~\eqref{eq:btagLR_s} is rigorous for
$\PW \to \PQu\PAQd(\PAQs)$ decays, whereas it is only approximate for $\PW\to \PQc\PAQs(\PAQd)$ decays,
since the CSV discriminator pdf for charm quarks differs with respect to $f_\mathrm{lf}$~\cite{Chatrchyan:2012jua}.
Equation~\eqref{eq:btagLR_s} can be extended to the case of SL events with five jets,
or DL events with at least four jets, by considering that in both cases four of the jets are associated with heavy-flavour partons,
and the remaining jets with light-flavour partons.
The likelihood under the alternative hypothesis, $f(\vec{\xi} | \ttpLF)$,
is given by Eq.~\eqref{eq:btagLR_s} after swapping $f_{\mathrm{hf}}$ for $f_{\mathrm{lf}}$.
The variable used to select events is then defined as the likelihood ratio
\begin{equation} \label{eq:btagLR}
\mathcal{F}( \vec{\xi} ) =  \frac{ f( \vec{\xi} | \ttpHF) } { f( \vec{\xi} | \ttpHF )+ f( \vec{\xi} | \ttpLF ) }.
\end{equation}
The distribution of $\mathcal{F}$ for SL events with six jets is shown in Fig.~\ref{fig:numJets} (bottom right).

In the following, events are retained if $\mathcal{F}$ is larger than a threshold value $\mathcal{F}_{\mathrm{L}}$
ranging between 0.85 and 0.97, depending on the channel and jet multiplicity.
The selected events are further classified as high-purity (low-purity) if $\mathcal{F}$ is larger
(smaller) than a value $\mathcal{F}_{\PH}$, with $\mathcal{F}_{\mathrm{L}}<\mathcal{F}_{\PH}<1.0$.
The low-purity categories serve as control regions for $\ttpLF$ jets,
providing constraints on several sources of systematic uncertainty.
The high-purity categories are enriched in $\ttpHF$ events, and drive the sensitivity of the analysis.
The thresholds $\mathcal{F}_{\mathrm{L}}$ and $\mathcal{F}_{\PH}$
are optimised separately for each of the analysis categories defined in Section~\ref{sec:algo}.
The exact values are reported in Table~\ref{tab:cutflow}.

After requiring a lower threshold on the selection variable $\mathcal{F}$,
the background is dominated by $\ttJJs$, with minor contributions from
the production of a single top quark plus jets, \ttbar plus vector bosons, and $\PW/\Z$+jets;
the expected purity for a SM Higgs boson signal is only at the percent level.
By construction, the selection criteria based on Eq.~\eqref{eq:btagLR} enhance the $\ttpBBbar$
subprocess compared to the otherwise dominant $\ttpLF$ production.
The $\ttpBBbar$ background has the same final state as the signal whenever the two \PQb quarks are resolved as individual jets.
Therefore, this background cannot be effectively reduced by means of the $\mathcal{F}$ discriminant.
The cross section for $\ttpBBbar$ production with two resolved \PQb-quark jets is larger than that of the
signal by about one order of magnitude and is affected by sizable theoretical uncertainties~\cite{ttbb},
which hampers the possibility of extracting the signal via a counting experiment.
A more refined approach,
which thoroughly uses the kinematic properties of the reconstructed event, is therefore required to
improve the separation between the signal and the background.

\section{Signal extraction} \label{sec:algo}

As in other resonance searches, the invariant mass reconstructed from the $\HBB$ decay
provides a natural discriminating variable to separate the narrow Higgs boson dijet resonance from the continuum mass
spectrum expected from the $\ttJJs$ background.
However, in the presence of additional \PQb quarks from the decay of the top quarks,
an ambiguity in the Higgs boson reconstruction is introduced, leading to a combinatorial background.
The distribution of the experimental mass estimator built from a randomly selected jet pair is much broader compared
to the detector resolution, since wrongly chosen jet pairs are only mildly or not at all correlated with $\mH$.
Unless a selection rule is introduced to filter out the wrong combinations,
the existence of such a combinatorial background results in a suppression of the statistical power of the mass estimator,
which grows as the factorial of the jet multiplicity.
Multivariate techniques that exploit the correlation between several observables in the same event are naturally
suited to deal with signal extraction in such complex final states.

In this paper, a likelihood technique based on the theoretical matrix elements for the $\ttH$
process and the $\ttpBBbar$ background is applied for signal extraction.
This method utilises the kinematics and dynamics of the event,
providing a powerful discriminant between the signal and background.
The $\ttpBBbar$ matrix elements are considered as the prototype to model all background processes.
This choice guarantees optimal separation between the signal and the $\ttpBBbar$ background,
which is a desirable property given the large rate and theoretical uncertainty in the latter.
The performance on the other $\ttJJs$ subprocesses might not be necessarily optimal,
even though some separation power is still preserved; indeed, the $\ttpBBbar$ matrix elements
describe these processes better than the signal matrix elements do, as it has been verified a posteriori with the simulation.
More specifically, the shapes of the matrix element discriminant predicted by the simulation for the different $\ttJJs$ subprocesses are found to be similar to each other, with a slightly better separation power for the $\ttpBBbar$ background. The approximate degeneracy in shape between several processes can be ascribed to a smearing effect of the combinatorial background, as well as to the impact of the Higgs boson mass constraint on the calculation of the event likelihood under the signal hypothesis. The latter provides a similar discrimination against all $\ttJJs$ subprocesses. A slightly worse separation power is instead observed for minor backgrounds, such as single top quark or $\ttZ$ events, for which neither of the two matrix elements tested really applies. However, all of the background processes analysed are found to yield discriminant shapes that can be well distinguished from that for the signal. Also, it is found that most of the statistical power attained by this method in separating $\ttH$, $\HBB$ from $\ttpBBbar$ events relies on the different correlation and kinematic distributions of the two \PQb-quark jets not associated with the top quark decays.

\subsection{Construction of the MEM probability density functions} \label{subsec:algo1}

The MEM probability density functions under the signal and background hypothesis are constructed at LO
assuming for simplicity that in both cases the reactions proceed via gluon fusion.
At $\sqrt{s}=8\TeV$, the fraction of the gluon-gluon initiated subprocesses is about 55\% (65\%) of the inclusive LO (NLO) cross section, and it grows with the centre-of-mass energy~\cite{Dawson:2003zu}.
Examples of diagrams entering the calculation are shown in the middle and right panels of Fig.~\ref{fig:fd}.
All possible jet-quark associations in the reconstruction of the final state are considered.
For each event, the MEM probability density function $w(\vec{y}|\mathcal{H})$
under the hypothesis $\mathcal{H}=\ttH$ or $\ttpBBbar$ is calculated as:
\ifthenelse{\boolean{cms@external}}{
\begin{multline} \label{eq:LM}
w(\vec{y}|\mathcal{H}) = \sum_{i=1}^{N_a} \int{ \frac{\rd{}x_{a} \rd{}x_{b}}{ 2x_{a}x_{b}s } }
\int{ \prod_{k=1}^{8} \left( \frac{\rd^3\vec{p}_{k}}{(2\pi)^32E_{k}} \right)}\times \\
\left( 2\pi \right)^4\delta^{(E,z)} \Bigl( p_{a}+p_{b} - \sum_{k=1}^{8} p_{k} \Bigr)
\: \mathcal{R}^{(x,y)} \Bigl( \vec{\rho}_{\mathrm{T}} , \sum_{k=1}^{8} p_{k} \Bigr) \times \\
g(x_{a}, \mu_{\mathrm{F}})g(x_{b}, \mu_{\mathrm{F}})
  \abs{\mathcal{M}_{\mathcal{H}} \left( p_{a}, p_{b}, p_1, \hdots, p_8\right)}^2 W\left( \vec{y} , \vec{p} \right),
\end{multline}
}{
\begin{equation} \label{eq:LM}
\begin{split}
w(\vec{y}|\mathcal{H}) = & \sum_{i=1}^{N_a} \int{ \frac{\rd{}x_{a} \rd{}x_{b}}{ 2x_{a}x_{b}s } }
\int{ \prod_{k=1}^{8} \left( \frac{\rd^3\vec{p}_{k}}{(2\pi)^32E_{k}} \right)}
\:\left( 2\pi \right)^4 \delta^{(E,z)} \Bigl( p_{a}+p_{b} - \sum_{k=1}^{8} p_{k} \Bigr)
\: \mathcal{R}^{(x,y)} \Bigl( \vec{\rho}_{\mathrm{T}} , \sum_{k=1}^{8} p_{k} \Bigr) \times \\
 &
g(x_{a}, \mu_{\mathrm{F}})g(x_{b}, \mu_{\mathrm{F}})
  \abs{\mathcal{M}_{\mathcal{H}} \left( p_{a}, p_{b}, p_1, \hdots, p_8\right)}^2 W\left( \vec{y} , \vec{p} \right),
\end{split}
\end{equation}
}
where $\vec{y}$ denotes the set of observables for which the matrix element pdf is constructed,
\ie the momenta of jets and leptons.
The sum extends over the $N_{a}$ possibilities of associating the jets with the final-state quarks.
The integration on the right-hand side of Eq.~\eqref{eq:LM} is performed over the phase space
of the final-state particles and over the gluon energy fractions $x_{a,b}$ by using the
\textsc{vegas}~\cite{VEGAS} algorithm.
The four-momenta of the initial-state gluons $p_{a,b}$ are related to the four-momenta of the colliding
protons $P_{a,b}$ by the relation $p_{a,b}=x_{a,b}P_{a,b}$.
The delta function
enforces the conservation of longitudinal momentum and energy between the incoming gluons and the $k=1,\hdots,8$
outgoing particles with four-momenta $p_{k}$.
To account for the possibility of inital/final state radiation, the total transverse momentum of the final-state particles, which should be identically zero at LO, is instead loosely constrained by the resolution function $\mathcal{R}^{(x,y)}$ to the measured transverse recoil $\vec{\rho}_{\mathrm{T}}$, defined
as the negative of the total transverse momentum of jets and leptons, plus the missing transverse momentum.

The remaining part of the integrand in Eq.~\eqref{eq:LM} contains the product of the gluon PDFs in the protons ($g$),
the square of the scattering amplitude ($\mathcal{M}$), and the transfer function ($W$).
For $\mathcal{H}=\ttH$, the factorisation scale $\mu_{\mathrm{F}}$ entering the PDF is taken as half of the sum of
twice the top-quark mass and the Higgs boson mass~\cite{Dawson:2002tg},
while for $\mathcal{H}=\ttpBBbar$ a dynamic scale is used equal to the quadratic sum of the transverse masses for
all coloured partons~\cite{Alwall:2003hc}. The scattering amplitude for the hard process is evaluated numerically at LO
accuracy by the program \textsc{OpenLoops}~\cite{OpenLoops};
all resonances are treated in the narrow-width approximation~\cite{NW}, and spin correlations are neglected.
The transfer function $W\left( \vec{y} , \vec{p} \right)$
provides a mapping between the measured set of observables $\vec{y}$ and the final-state particles momenta $ \vec{p}=(\vec{p}_1, \hdots, \vec{p}_8)$.
Given the good angular resolution of jets, the direction of quarks is assumed to be perfectly measured by the direction of the associated jets. Also, since energies of leptons are measured more precisely than for jets, their momenta are considered perfectly measured.
Under these assumptions, the total transfer function reduces to the product of the quark energy transfer function
times the probability for the quarks that are not reconstructed as jets to fail the acceptance criteria.
The quark energy transfer function is modelled by a single Gaussian function for jets associated with light-flavour partons,
and by a double Gaussian function for jets associated with bottom quarks; the latter are constructed by superimposing two Gaussian functions with different mean and standard deviation. Such an asymmetric parametrisation provides a good description of both the core of the detector energy response and the low-energy tail arising from semileptonic B hadron decays.
The parametrisation of the transfer functions has been derived from MC simulated samples.

\subsection{Event categorisation} \label{subsec:algo2}

To aid the evaluation of the MEM probability density functions at LO,
events are classified into mutually exclusive categories based on different parton-level interpretations.
Firstly, the set of jets yielding the largest contribution to the sum
defined by Eq.~\eqref{eq:btagLR_s}, determines the four (tagged) jets associated with bottom
quarks; the remaining $N_\text{untag}$ (untagged) jets
are assumed to originate either from $\PW\to\qqbar'$ decays (SL channel) or from initial- or final-state gluon radiation (SL and DL channels).
There still remains a twelve-fold ambiguity in the determination of the parton matched to each jet, which is reflected by
the sum in Eq.~\eqref{eq:LM}.
Indeed, without distinguishing between \PQb and \PAQb quarks, there exist
$4!/(2!\:2!) = 6$ combinations for assigning two jets out of four with the
Higgs boson decay ($\mathcal{H}=\ttH$), or with the bottom quark-pair radiation ($\mathcal{H}=\ttpBBbar$);
for each of these possibilities,
there are two more ways of assigning the remaining tagged jets to either the \PQt or \PAQt quark,
thus giving a total of twelve associations.
In the SL channel, an event can be classified in one of three possible categories.
The first category (Cat-1) is defined by requiring at least six jets; if there are exactly six jets, the mass of the two untagged jets is required to
be in the range $\left[60,100\right]$\GeV, \ie compatible with the mass of the $\PW$ boson.
If the number of jets is larger than six, the mass range is tightened
to compensate for the increased ambiguity in selecting the correct \PW~boson decay products.
In the event interpretation, the $\PW\to\qqbar'$ decay is assumed to be fully reconstructed,
with the two quarks identified with the jet pair satisfying the mass constraint.
The definition of the second category (Cat-2) differs from that of Cat-1 by the inversion of the dijet mass constraint.
This time, the event interpretation assumes that one of the quarks from the $\PW$ boson decay has failed the reconstruction.
The integration on the right-hand side of Eq.~\eqref{eq:LM} is extended to include the phase space of the nonreconstructed quark.
The other untagged jet(s) is (are) interpreted as gluon radiation, and do not enter the calculation of $w(\vec{y}|\mathcal{H})$.
The total number of associations considered is twelve times the multiplicity of untagged jets
eligible to originate from the $\PW$ boson decay: $N_{a}=12 N_\text{untag}$.
In the third category (Cat-3), exactly five jets are required, and an incomplete $\PW$ boson reconstruction is again assumed.
In the DL channel, only one event interpretation is considered, namely that each of the four bottom quarks in the decay is associated with one of the four tagged jets.

Finally, two event discriminants, denoted by $P_{\mathrm{s/b}}$ and $P_{\mathrm{h/l}}$, are defined.
The former encodes only information from the event kinematics and dynamics via Eq.~\eqref{eq:LM},
and is therefore suited to separate the signal from the background;
the latter contains only information related to \PQb tagging,
thus providing a handle to distinguish between the heavy- and the light-flavour components of the $\ttJJs$ background. They are defined as follows:
\ifthenelse{\boolean{cms@external}}{
\begin{align}\label{eq:P_sb}
P_{\mathrm{s/b}} &= \frac{ w(\vec{y} | \ttH ) }{ w(\vec{y} | \ttH) +  k_{\mathrm{s/b}}\, w(\vec{y} | \ttpBBbar) }\\
\intertext{and}
P_{\mathrm{h/l}} &= \frac{ f(  \vec{\xi} | \ttpHF) }{ f(  \vec{\xi} | \ttpHF) + k_{\mathrm{h/l}}\, f(  \vec{\xi} | \ttpLF)   },
\end{align}
}{
\begin{equation}\label{eq:P_sb}
P_{\mathrm{s/b}} = \frac{ w(\vec{y} | \ttH ) }{ w(\vec{y} | \ttH) +  k_{\mathrm{s/b}} w(\vec{y} | \ttpBBbar) } \quad\text{and}\quad
P_{\mathrm{h/l}} = \frac{ f(  \vec{\xi} | \ttpHF) }{ f(  \vec{\xi} | \ttpHF) + k_{\mathrm{h/l}} f(  \vec{\xi} | \ttpLF)   },
\end{equation}
}
where the functions $f( \vec{\xi} | \ttpHF )$ and $f( \vec{\xi} | \ttpLF)$ are defined
as in Eq.~\eqref{eq:btagLR_s} but restricting the sum only to the jet-quark associations considered in the calculation of $w(\vec{y})$;
the coefficients $k_{\mathrm{s/b}}$ and $k_{\mathrm{h/l}}$ in the denominators are positive constants that can differ among the categories and will be treated as optimisation parameters, as described below.

The joint distribution of the $(P_{\mathrm{s/b}},P_{\mathrm{h/l}})$ discriminants is used in
a two-dimensional maximum likelihood fit to search for events resulting
from Higgs boson production. By construction, the two discriminants satisfy
the constraint $0 \leq P_{\mathrm{s/b}}, P_{\mathrm{h/l}} \leq 1$.
Because of the limited size of the simulated samples,
the distributions of $P_{\mathrm{s/b}}$ and $P_{\mathrm{h/l}}$ are binned.
A finer binning is used for the former, which carries the largest sensitivity to the signal,
while the latter is divided into two equal-sized bins.
The coefficient $k_{\mathrm{s/b}}$ appearing in the definition of $P_{\mathrm{s/b}}$ is introduced to adjust the relative normalisation
between $w( \vec{y}  | \ttH )$ and $w(\vec{y}  | \ttpBBbar )$; likewise for $k_{\mathrm{h/l}}$.
A redefinition of any of the two coefficients would change the corresponding discriminant monotonically,
thus with no impact on its separation power.
However, since both variables are analysed in bins with fixed size, an optimisation procedure, based on minimising the
expected exclusion limit on the signal strength as described in Section~\ref{sec:results},
is carried out to choose the values that maximise the sensitivity of the analysis.
More specifically, the coefficients $k_{\mathrm{s/b}}$ are first set to the values that remove any local maximum for the $\ttpBBbar$ distribution around $P_{\mathrm{s/b}}\sim1$, a condition that is found to provide already close to optimal coefficients. Then, starting from this initial point, several values of $k_{\mathrm{s/b}}$ are scanned and the $P_{\mathrm{s/b}}$ distributions are recomputed accordingly. An expected upper limit on the signal strength is then evaluated for each choice of $k_{\mathrm{s/b}}$
using the simulated samples. This procedure is repeated until a minimum in the expected limit is obtained.
A similar procedure is applied for choosing the optimal $k_{\mathrm{h/l}}$ coefficients.

\subsection{Background modelling} \label{subsec:algo3}

The background normalisation and the distributions of the event discriminants
are derived by using the MC simulated samples described in Section~\ref{sec:data}.
In light of the large theoretical uncertainty that affects the prediction of \ttbar plus heavy-flavour~\cite{ttbb,ttbb_old},
the \MADGRAPH sample is further divided into subsamples based on the quark flavour associated
with the jets generated in the acceptance region $\pt>20\GeV$, $\abs{\eta}<2.5$.
Events are labelled as $\ttpBBbar$ if at least two jets are matched within $\sqrt{\smash[b]{(\Delta\eta)^2+(\Delta\phi)^2}}<0.5$
to bottom quarks not originating from the decay of a top quark. If only one jet is matched to a bottom quark, the event is labelled as
$\ttpB$. These cases typically arise when the second extra \PQb quark in the event is either too far
forward or too soft to be reconstructed as a jet, or because the two extra \PQb quarks are emitted almost collinearly and end up in a single jet. Similarly, if at least one reconstructed jet is matched to a \PQc quark, the event is labelled as $\ttpCCbar$. In the latter case, single- and double-matched events are treated as one background. If none of the above conditions is satisfied, the event is classified as \ttbar plus light-flavour.
Table~\ref{tab:cutflow} reports the number of events observed in the various categories,
together with the expected signal and background yields.
The latter are obtained from the signal-plus-background fit described in Section~\ref{sec:results}.

\begin{table*}[htbp]
\centering
\topcaption{Expected and observed event yields in the (top) high-purity (H) and (bottom) low-purity (L) categories of the SL and DL channels.
The expected event yields with their uncertainties are obtained from a
signal-plus-background fit as described in Section~\ref{sec:results}.
In the last row of each table, the symbol S (B) denotes the signal (total background) yield.}
\ifthenelse{\boolean{cms@external}}{}{\resizebox{\textwidth}{!}}
{
\begin{tabular}{ *{5}{c} }
 & SL Cat-1 (H)  & SL Cat-2 (H)  & SL Cat-3 (H)  & DL (H) \\
 \PQb-tagging selection & $0.995\leq\mathcal{F}\leq1.000$  & $0.993\leq\mathcal{F}\leq1.000$ & $0.995\leq\mathcal{F}\leq1.000$ & $0.925\leq\mathcal{F}\leq1.000$ \\ \hline
 $\ttH$  &  3.5 $\pm$ 0.7  &  5.9 $\pm$ 1.1  &  7.5 $\pm$ 1.4  &  4.5 $\pm$ 0.7 \\
\hline
 $\ttbar\mathrm{+lf}$  &  17 $\pm$ 3  &  70 $\pm$ 13  &  152 $\pm$ 21  &  84 $\pm$ 11 \\
 $\ttbar\mathrm{+}\ccbar$  &  22 $\pm$ 8  &  66 $\pm$ 20  &  81 $\pm$ 24  &  85 $\pm$ 24 \\
 $\ttbar\mathrm{+}\PQb$  &  16 $\pm$ 8  &  44 $\pm$ 23  &  70 $\pm$ 32  &  47 $\pm$ 23 \\
 $\ttbar\mathrm{+}\bbbar$   &  43 $\pm$ 11  &  75 $\pm$ 17  &  69 $\pm$ 18  &  50 $\pm$ 13 \\
 $\ttbar\mathrm{+}\PW/\Z$  &  3.2 $\pm$ 0.8  &  4.4 $\pm$ 1.1  &  4.2 $\pm$ 1.0  &  5.1 $\pm$ 1.1 \\
Single $\PQt$  &  3.1 $\pm$ 1.4  &  5.3 $\pm$ 2.2  &  14 $\pm$ 4  &  5.9 $\pm$ 1.7 \\
$\PW/\Z$+jets  &  \NA  &  0.3 $\pm$ 2.2  &  \NA  &  6 $\pm$ 5 \\
\hline
 Total background  & 103 $\pm$ 11   &  265 $\pm$ 24   &  390 $\pm$ 28   & 283 $\pm$ 24  \\
\hline
 Data  &  107  &  272  &  401  &  279 \\
\hline
$ \mathrm{S/B}$ ($\mathrm{S/\sqrt{B}}$)  & 3.4\% (0.34)  & 2.2\% (0.36)  & 1.9\% (0.38)  & 1.6\% (0.27) \\
\end{tabular}
}
\mbox{}\\[1ex]
\hrule
\mbox{}\\[0.5ex]
\hrule
\mbox{}\\[1ex]
\ifthenelse{\boolean{cms@external}}{}{\resizebox{\textwidth}{!}}
{
\begin{tabular}{ *{5}{c}  }
 & SL Cat-1 (L)  & SL Cat-2 (L)  & SL Cat-3 (L)  & DL (L) \\
  \PQb-tagging selection & $0.960\leq\mathcal{F}<0.995$ &  $0.960\leq\mathcal{F}<0.993$ & $0.970\leq\mathcal{F}<0.995$ & $0.85\leq\mathcal{F}<0.925$ \\ \hline
 $\ttH$  &  3.8 $\pm$ 0.7  &  5.2 $\pm$ 0.8  &  7.9 $\pm$ 1.3  &  0.8 $\pm$ 0.1 \\
\hline
 $\ttbar\mathrm{+lf}$  &  111 $\pm$ 13  &  268 $\pm$ 32  &  737 $\pm$ 62  &  69 $\pm$ 8 \\
 $\ttbar\mathrm{+}\ccbar$  &  94 $\pm$ 25  &  161 $\pm$ 45  &  268 $\pm$ 74  &  40 $\pm$ 11 \\
 $\ttbar\mathrm{+}\PQb$  &  45 $\pm$ 24  &  80 $\pm$ 42  &  162 $\pm$ 77  &  14 $\pm$ 7 \\
 $\ttbar\mathrm{+}\bbbar$   &  48 $\pm$ 13  &  69 $\pm$ 17  &  84 $\pm$ 22  &  7.6 $\pm$ 2.1 \\
 $\ttbar\mathrm{+}\PW/\Z$  &  4.0 $\pm$ 1.0  &  6.9 $\pm$ 1.5  &  7.8 $\pm$ 1.6  &  2.3 $\pm$ 0.5 \\
Single $\PQt$  &  8.9 $\pm$ 2.4  &  13 $\pm$ 3  &  32 $\pm$ 6  &  3.1 $\pm$ 1.1 \\
$\PW/\Z$+jets  &  \NA  &  \NA  &  \NA  &  5 $\pm$ 3 \\
\hline
 Total background  &  311 $\pm$ 22  &  598 $\pm$ 38   & 1291 $\pm$ 60   &  142 $\pm$ 10 \\
\hline
 Data  &  310  &  603  &  1310  &  140 \\
\hline
$ \mathrm{S/B}$ ($\mathrm{S/\sqrt{B}}$)  & 1.2\% (0.21)  & 0.9\% (0.21)  & 0.6\% (0.22)  & 0.5\% (0.07) \\

\end{tabular}
\label{tab:cutflow}
}
\end{table*}

\section{Systematic uncertainties} \label{sec:systematics}

There are a number of systematic uncertainties of experimental and theoretical origin that affect the signal and the background expectations.
Each source of systematic uncertainty is associated with a nuisance parameter that modifies the likelihood function used to extract
the signal yield, as described in Section~\ref{sec:results}. The prior knowledge on the nuisance parameter is incorporated into the likelihood
in a frequentist manner by interpreting it as a posterior arising from a pseudo-measurement~\cite{Combine}.
Nuisance parameters can affect either the yield of a process (normalisation uncertainty),
or the shape of the $P_{\mathrm{s/b}}$ and $P_{\mathrm{h/l}}$ discriminants (shape uncertainty),
or both. Multiple processes across several categories can be affected by the same source of uncertainty.
In that case the related nuisance parameters are treated as fully correlated.

The uncertainty in the integrated luminosity is estimated to be 2.6\%~\cite{CMS-PAS-LUM-13-001}.
The lepton trigger, reconstruction, and identification efficiencies are determined from control regions by
using a tag-and-probe procedure. The total uncertainty is evaluated from the statistical uncertainty of the tag-and-probe measurement,
plus a systematic uncertainty in the method, and is estimated to be 1.6\% per muon and 1.5\% per electron.
It is conservatively approximated to a constant 2\% per charged lepton.
The uncertainty on the jet energy scale (JES) ranges from 1\% up to about 8\% of the expected energy scale
depending on the jet $\pt$ and $\abs{\eta}$~\cite{cmsJEC}.
For each simulated sample, two alternative distributions of the $P_{\mathrm{s/b}}$ and $P_{\mathrm{h/l}}$
discriminants are obtained by varying the energy scale of all simulated jets up or down by their uncertainty,
and the fit is allowed to interpolate between the nominal and the alternative distributions with a Gaussian prior~\cite{Combine}.
A similar procedure is applied to account for the uncertainty related to the jet energy resolution (JER), which ranges between about
5\% and 10\% of the expected energy resolution depending on the jet direction. Since the analysis categories are defined in terms of the multiplicity
and kinematic properties of the jets, a variation of either the scale or the resolution of the simulated jets can induce a migration
of events in or out of the analysis categories, as well as migrations among different categories.
The fractional change in the event yield induced by a shift of the JES (JER)
ranges between 4--13\% (0.5--2\%), depending on the process type and on the category.
When the JES and JER are varied from their nominal values, the $\ptvecmiss$ vector is recomputed accordingly.
The scale factors applied to correct the CSV discriminator, as described in Section~\ref{sec:obj},
are affected by several sources of systematic uncertainty.
In the statistical interpretation, the fit can interpolate between the nominal and the two alternative distributions
constructed by varying each scale factor up or down by its uncertainty.

Theoretical uncertainties are treated as process-specific if they impact
the prediction of one simulated sample at the time.
They are instead treated as correlated across several samples if they are related
to common aspects of the simulation (\eg PDF, scale variations).
The modelling of the $\ttJJs$ background is affected by a variety of systematic uncertainties.
The uncertainty due to the top-quark \pt modelling is evaluated by varying the reweighting function
$r_{\PQt}(\pt^{\PQt})$, where $\pt^{\PQt}$ is the transverse momentum of the generated top quark, between
one (no correction at all) and $2r_{\PQt}-1$ (the relative correction is doubled). This results in both a shape and a normalisation uncertainty. The latter can be as large as 20\% for a top quark \pt around 300\GeV, and corresponds
to an overall normalisation uncertainty of about 3--8\% depending on the category.
To account for uncertainties in the $\ttJJs$ acceptance,
the factorisation and renormalisation scales used in the simulation are varied in a correlated way
by factors of $1/2$ and 2 around their central value.
The scale variation is assumed uncorrelated among $\ttpBBbar$, $\ttpB$, and
$\ttpCCbar$. In a similar way, independent scale variations are introduced for events with exactly one, two, or three extra partons
in the matrix element.
To account for possibly large $\mathrm{K}$-factors due to the usage of a LO MC
generator, the $\ttpBBbar$, $\ttpB$, and $\ttpCCbar$ normalisations predicted by the \MADGRAPH simulation
are assigned a 50\% uncertainty each. This value can be seen as a conservative upper limit to the theoretical uncertainty
in the $\ttpHF$ cross section achieved to date~\cite{ttbb}. Essentially, the approach followed here is to assign
large a priori normalisation uncertainties to the different $\ttJJs$ subprocesses, thus allowing the fit to simultaneously
adjust their rates.
Scale uncertainties in the inclusive theoretical cross sections used to normalise the simulated samples
range from a few percent up to 20\%, depending on the process.
The PDF uncertainty is treated as fully correlated for all processes that share the same dominant initial state
(\ie $\Pg\Pg$, $\Pg\PQq$, or $\PQq\PQq$); it ranges between 3\% and 9\%,
depending on the process.
Finally, the effect of the limited size of the simulated samples is accounted for by introducing one nuisance parameter for each
bin of the discriminant histograms and for each sample, as described in Ref.~\cite{BarlowBeeston}.
Table~\ref{table:syst} summarises the various sources of systematic uncertainty with their impact on the analysis.

\begin{table*}[htbp]
\topcaption{
Summary of the systematic uncertainties affecting the signal and background expectation.
The second column reports the range of rate variation for the processes affected by a given source of systematic uncertainty (as specified in the last three columns) when the nuisance parameter associated with it is varied up or down by its uncertainty.
The third column indicates whether a source of systematic uncertainty is assumed to affect the process normalisation only,
or both the normalisation and the shape of the event discriminants.
}
\centering
\begin{tabular}{ *{6}{c}}
                                      \multirow{2}{*}{Source}    &  \multirow{2}{*}{Rate uncertainty}   &  \multirow{2}{*}{Shape}   &  \multicolumn{3}{ c }{Process}    \\ \cline{4-6}
                                                                             &                                            &                                      &\rule[0pt]{0pt}{2.3ex} $\ttH$ & $\ttJJs$ & Others\\

\hline
\multicolumn{6}{ c }{Experimental uncertainties}   \\
\hline
  Integrated luminosity                     & 2.6\% & No & \checkmark & \checkmark & \checkmark \\
  Trigger and lepton identification    & 2--4\% & No & \checkmark & \checkmark & \checkmark\\
  JES                                                 &    4--13\%        & Yes & \checkmark & \checkmark & \checkmark\\
  JER                                                 &     0.5--2\%       & Yes & \checkmark & \checkmark & \checkmark\\
  \PQb tagging                                       &   2--17\%         & Yes & \checkmark & \checkmark & \checkmark\\

\hline
\multicolumn{6}{ c }{Theoretical uncertainties}   \\
\hline

  Top $\pt$ modelling             &   3--8\%        & Yes &  & \checkmark &  \\
  $\mu_{\mathrm{R}}/\mu_{\mathrm{F}}$ variations     &    2--25\%       & Yes &   & \checkmark &  \\
   $\ttpBBbar$ normalisation            &     50\%      & No &  & \checkmark &  \\
   $\ttpB$ normalisation             &      50\%     & No &  & \checkmark &  \\
  $\ttpCCbar$ normalisation            &      50\%     & No &  & \checkmark &  \\
  Signal cross section     &    7\%       & No &  \checkmark &  &  \\
  Background cross sections     &   2--20\%        & No &   & \checkmark & \checkmark \\
  PDF                     &  3--9\%         & No & \checkmark & \checkmark & \checkmark \\
  Statistical uncertainty (bin-by-bin)            &     4--30\%      & Yes & \checkmark & \checkmark & \checkmark\\
\end{tabular}
\label{table:syst}
\end{table*}
\section{Results} \label{sec:results}

The statistical interpretation of the results is performed by using the same methodology employed for other CMS Higgs boson analyses
and extensively documented in Ref.~\cite{Higgs1}.
The measured signal rate is characterised by a strength modifier $\mu=\sigma/\sigma_{\mathrm{SM}}$
that scales the Higgs boson production cross section times branching fraction
with respect to its SM expectation for $\mH=125$\GeV. The nuisance parameters, $\theta$, are incorporated into the likelihood as described in Section~\ref{sec:systematics}.
The total likelihood function $\mathcal{L}\left(\mu,\theta\right)$ is the product of a Poissonian likelihood spanning all bins of the $(P_{\mathrm{s/b}},P_{\mathrm{h/l}})$ distributions for all the eight categories, times a likelihood function for the nuisance parameters.
Based on the asymptotic properties of the profile likelihood ratio test statistic $q(\mu) = -2\ln\bigl[\mathcal{L}\bigl(\mu,\hat{\theta}_{\mu}\bigr)/\mathcal{L}\bigl(\hat{\mu},\hat{\theta}\bigr)\bigr]$, confidence intervals on $\mu$ are set,
where $\hat{\theta}$ and $\hat{\theta}_{\mu}$ indicate the best-fit value for $\theta$
obtained when $\mu$ is floating in the fit or fixed at a hypothesised value, respectively.

\begin{figure*}[htbp]
\centering
  \includegraphics[width=0.475\linewidth]{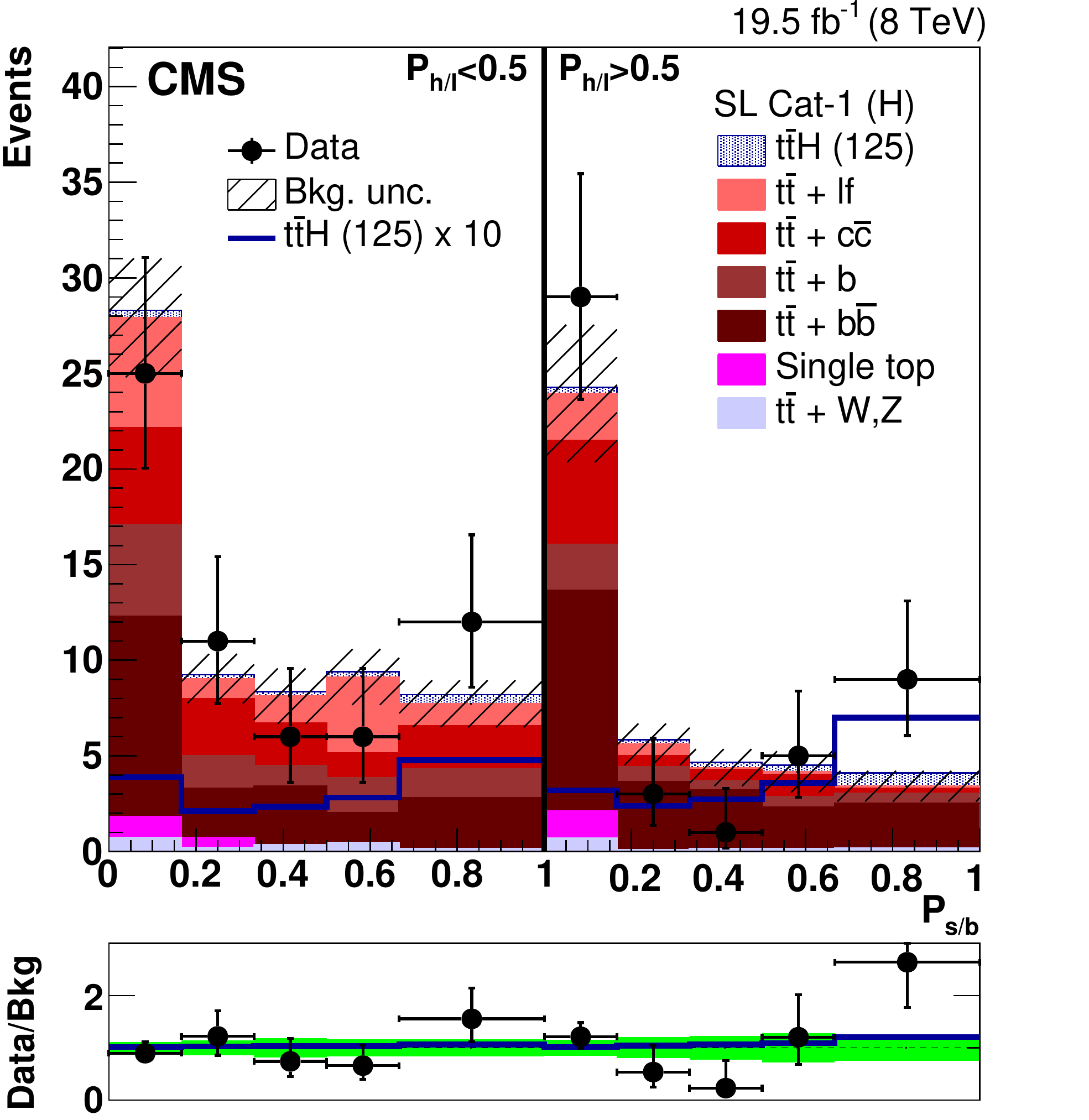}
  \includegraphics[width=0.475\linewidth]{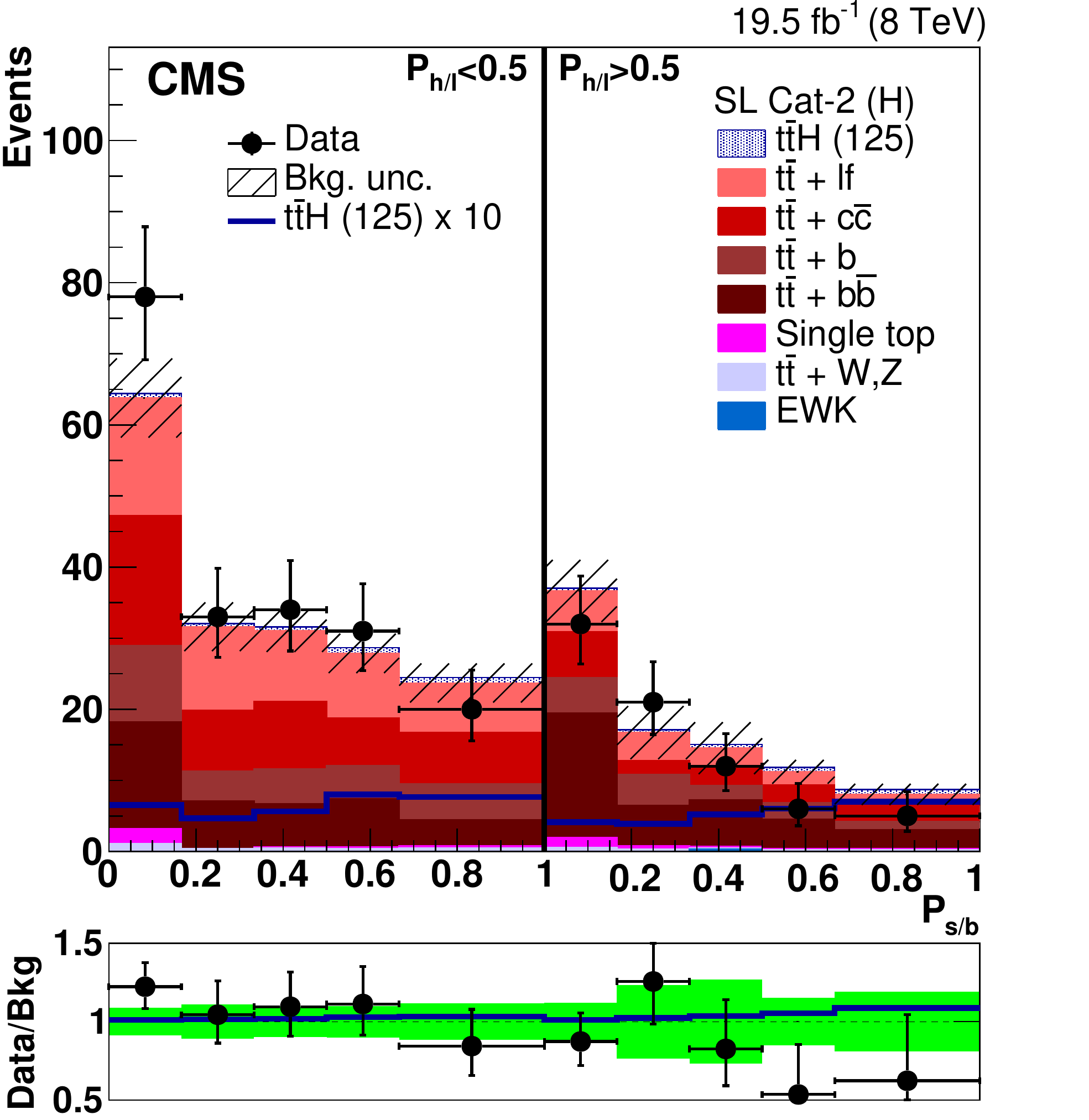}\\
  \includegraphics[width=0.475\linewidth]{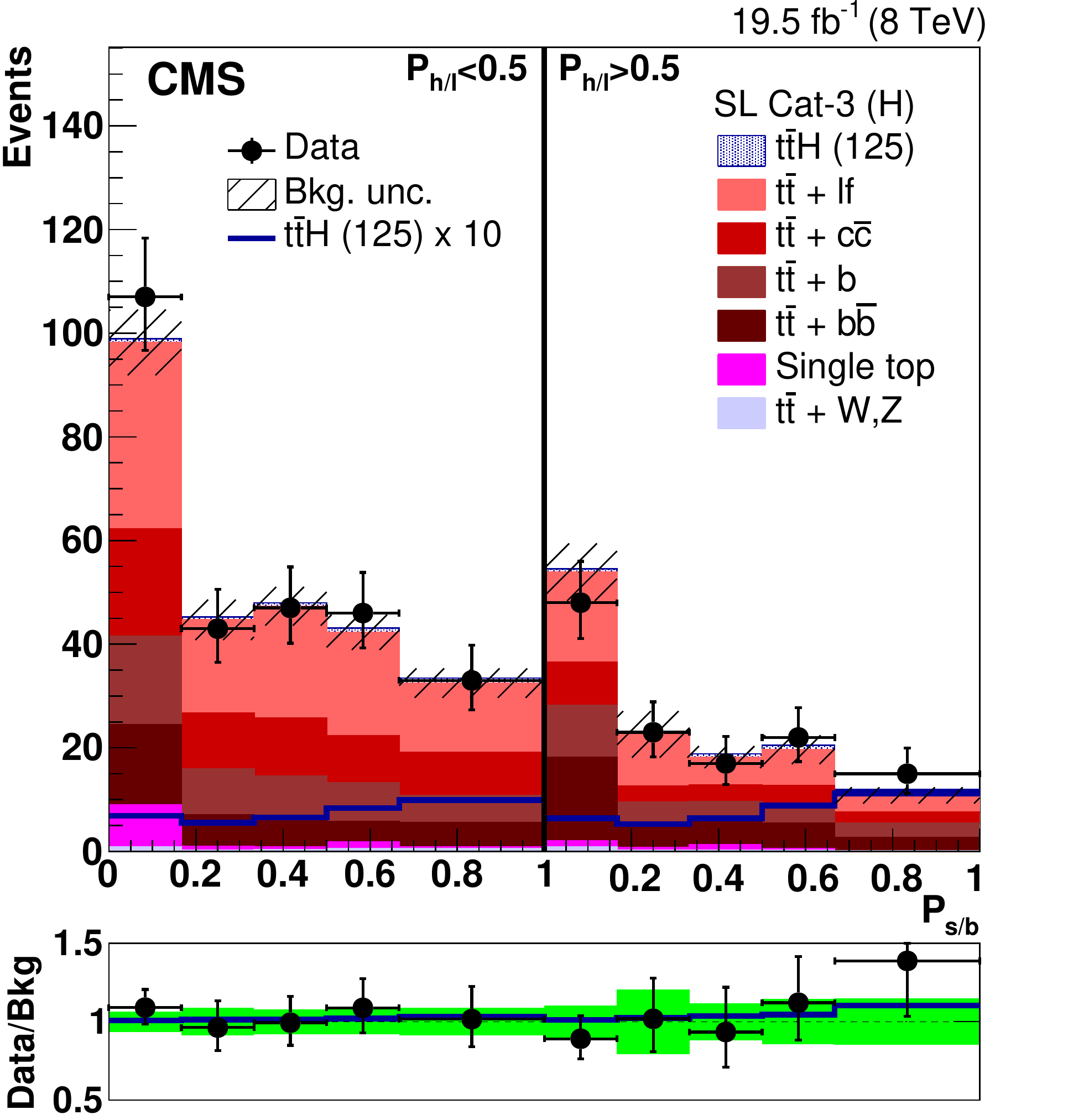}
  \includegraphics[width=0.475\linewidth]{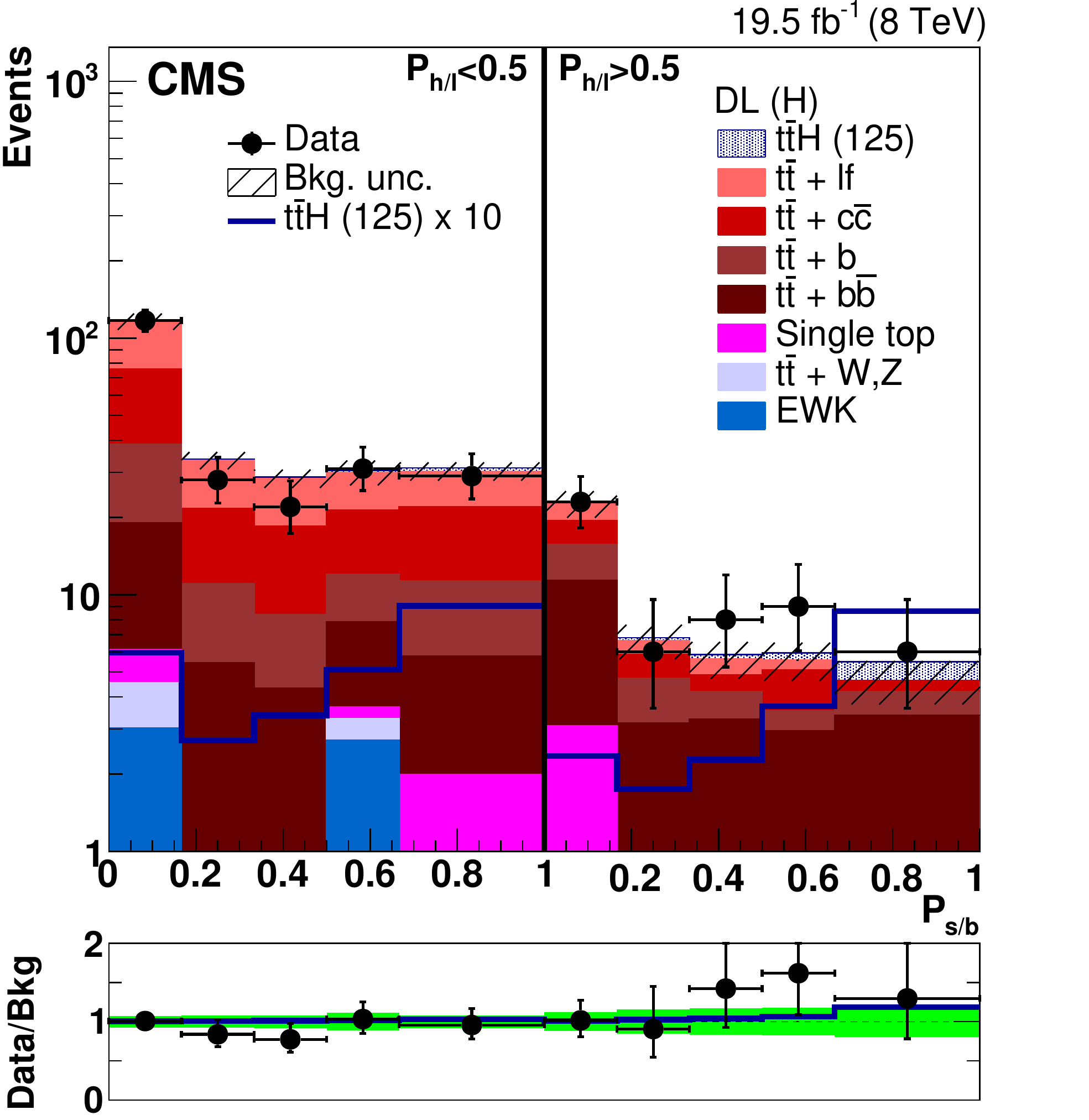}\\
\caption{
Distribution of the $P_{\mathrm{s/b}}$ discriminant in the two $P_{\mathrm{h/l}}$ bins
for the high-purity (H) categories.
The signal and background yields have been obtained from a combined fit of all nuisance parameters with the constraint $\mu=1$.
The bottom panel of each plot shows the ratio between the observed and the overall background yields.
The solid blue line indicates the ratio between the signal-plus-background and the background-only distributions.
The shaded and solid green bands band correspond to the $\pm1\sigma$ uncertainty in the background prediction after the fit.
}
\label{fig:cat_post_all_H}
\end{figure*}

\begin{figure*}[htbp]
\centering
  \includegraphics[width=0.475\linewidth]{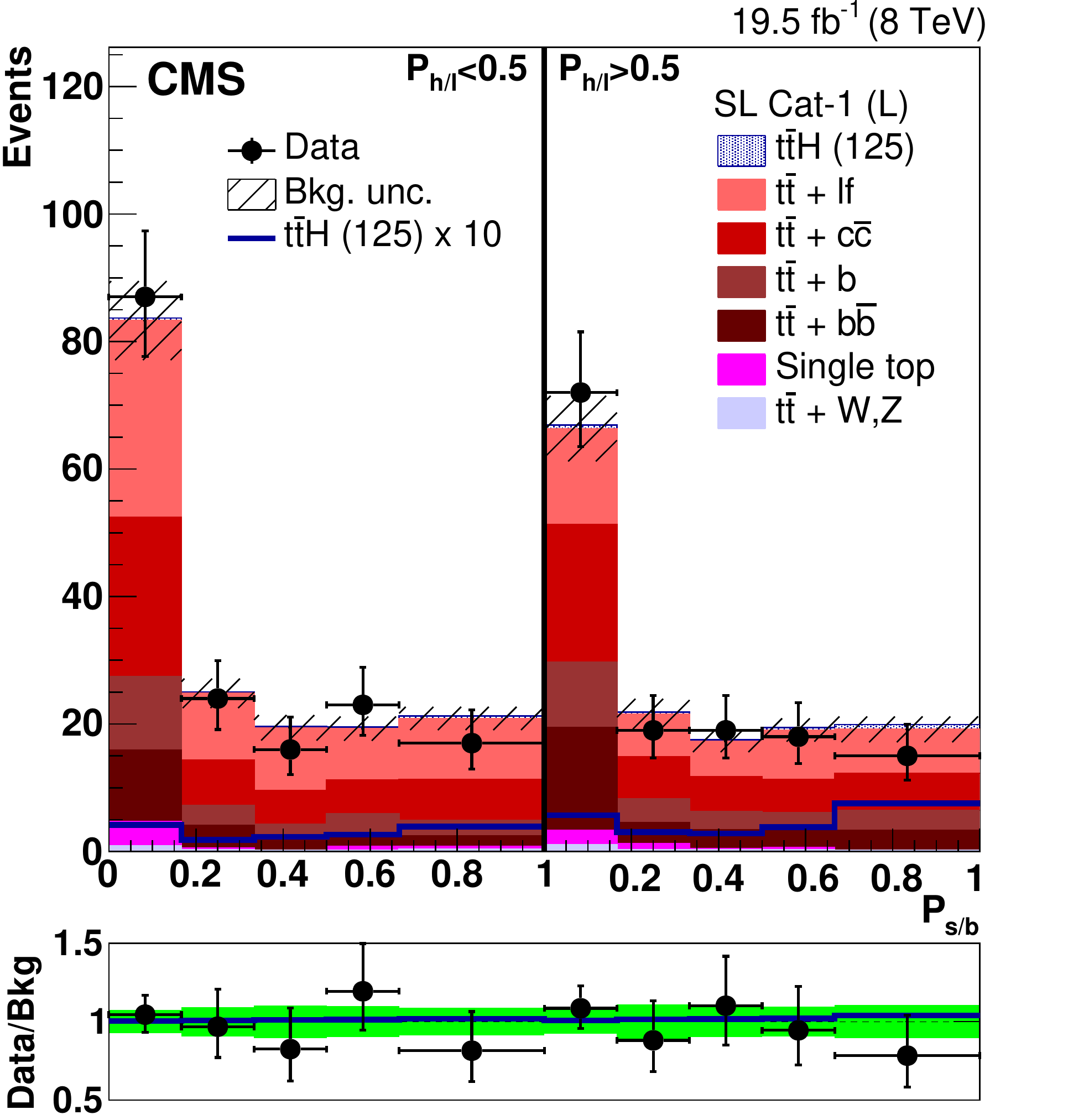}
  \includegraphics[width=0.475\linewidth]{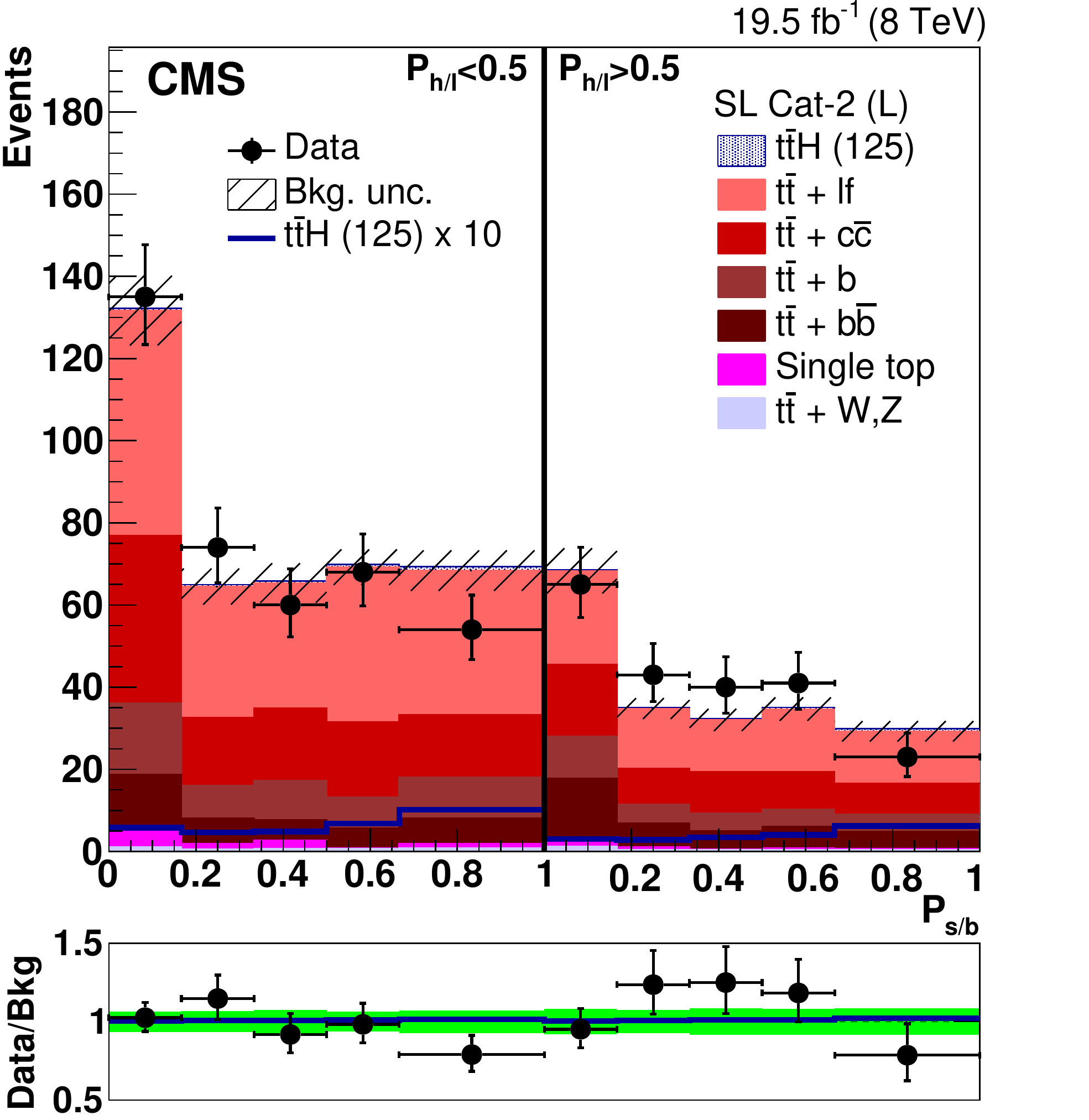}\\
  \includegraphics[width=0.475\linewidth]{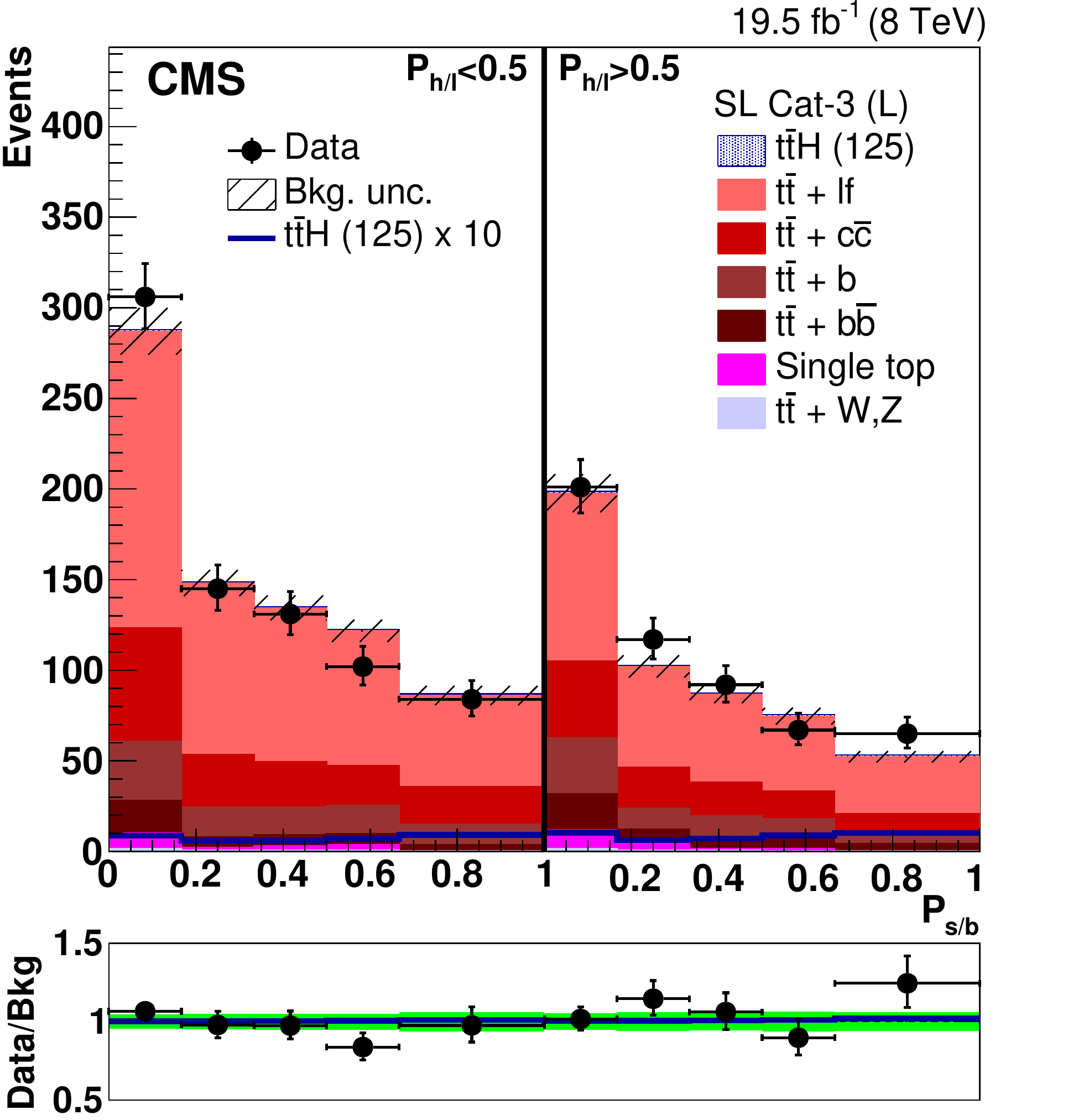}
  \includegraphics[width=0.475\linewidth]{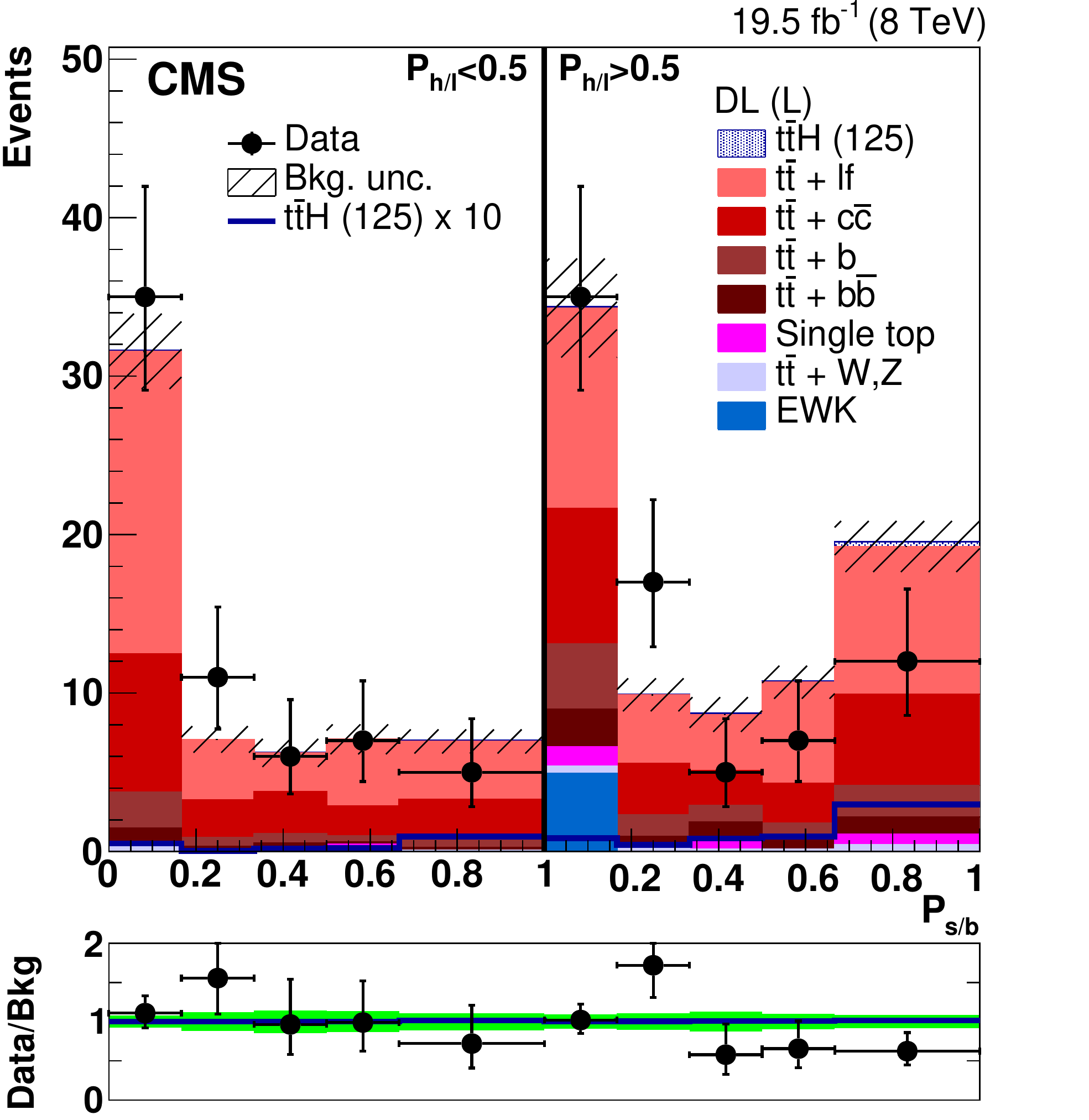}\\
\caption{
Distribution of the $P_{\mathrm{s/b}}$ discriminant in the two $P_{\mathrm{h/l}}$ bins
for the low-purity (L) categories.
The signal and background yields have been obtained from a combined fit of all nuisance parameters with the constraint $\mu=1$.
The bottom panel of each plot shows the ratio between the observed and the overall background yields.
The solid blue line indicates the ratio between the signal-plus-background and the background-only distributions.
The shaded and solid green bands correspond to the $\pm1\sigma$ uncertainty in the background prediction after the fit.
}
\label{fig:cat_post_all_L}
\end{figure*}

{\tolerance=800
Figures~\ref{fig:cat_post_all_H} and \ref{fig:cat_post_all_L} show the binned distributions of $(P_{\mathrm{s/b}}, P_{\mathrm{h/l}})$ in the various categories and for the two channels.
For visualisation purposes, the two-dimensional histograms are projected onto one dimension by showing first the distribution of $P_{\mathrm{s/b}}$ for events with $P_{\mathrm{h/l}}<0.5$ and then for $P_{\mathrm{h/l}}\geq0.5$.
The observed distributions are compared to the signal-plus-background expectation obtained from a combined fit to all categories
with the constraint $\mu=1$. No evidence of a $\ttH$ signal over the background is observed.
The statistical interpretation is performed both in terms of exclusion upper limits (UL) at a 95\% \CL,
where the modified \CLs prescription \cite{Read1,junkcls} is adopted to quote confidence intervals,
and in terms of the maximum likelihood estimator of the strength modifier ($\hat{\mu}$).
\par}

\begin{figure}[htbp]
\centering
  \includegraphics[width=0.475\textwidth]{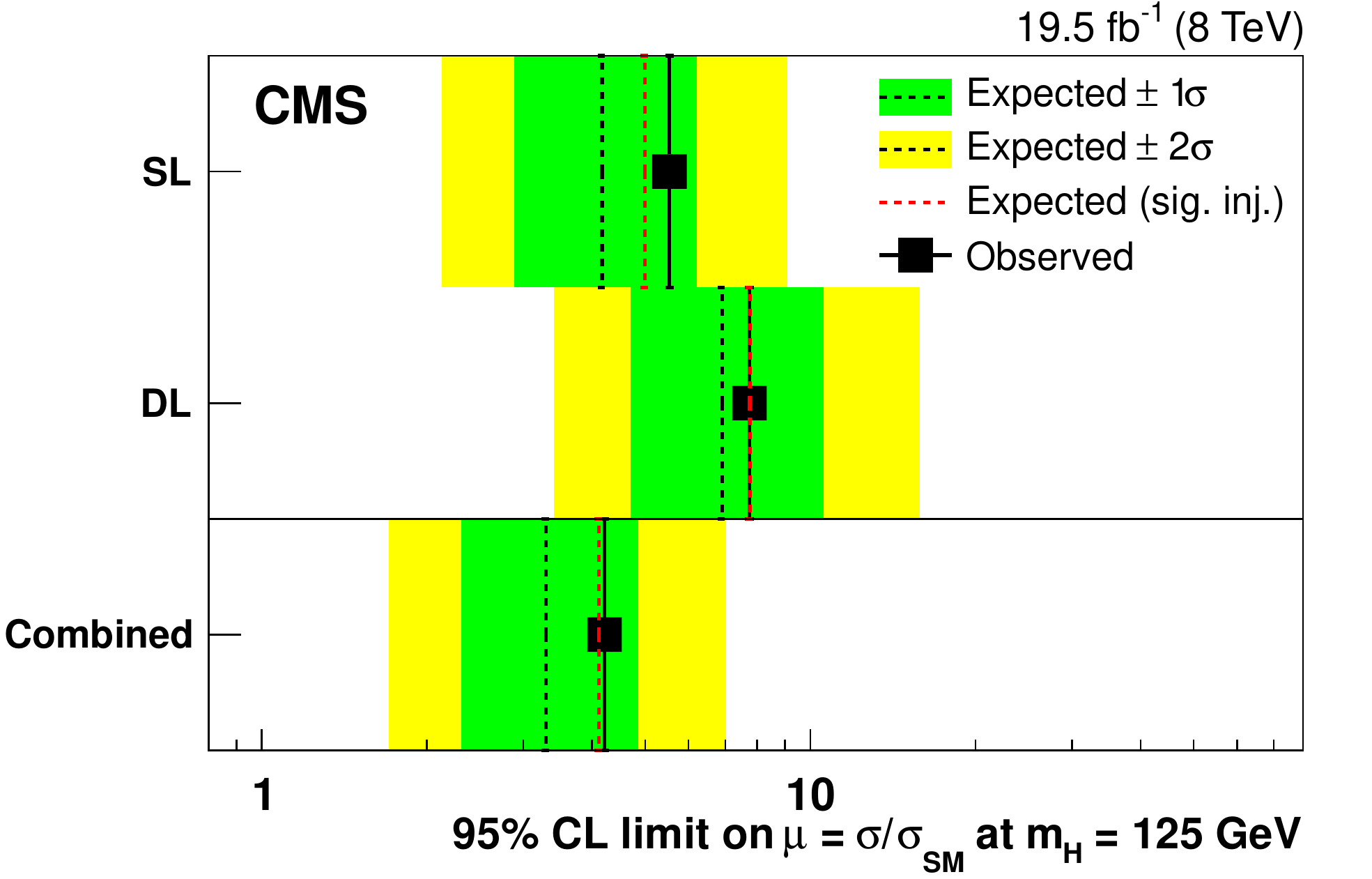}
  \includegraphics[width=0.475\textwidth]{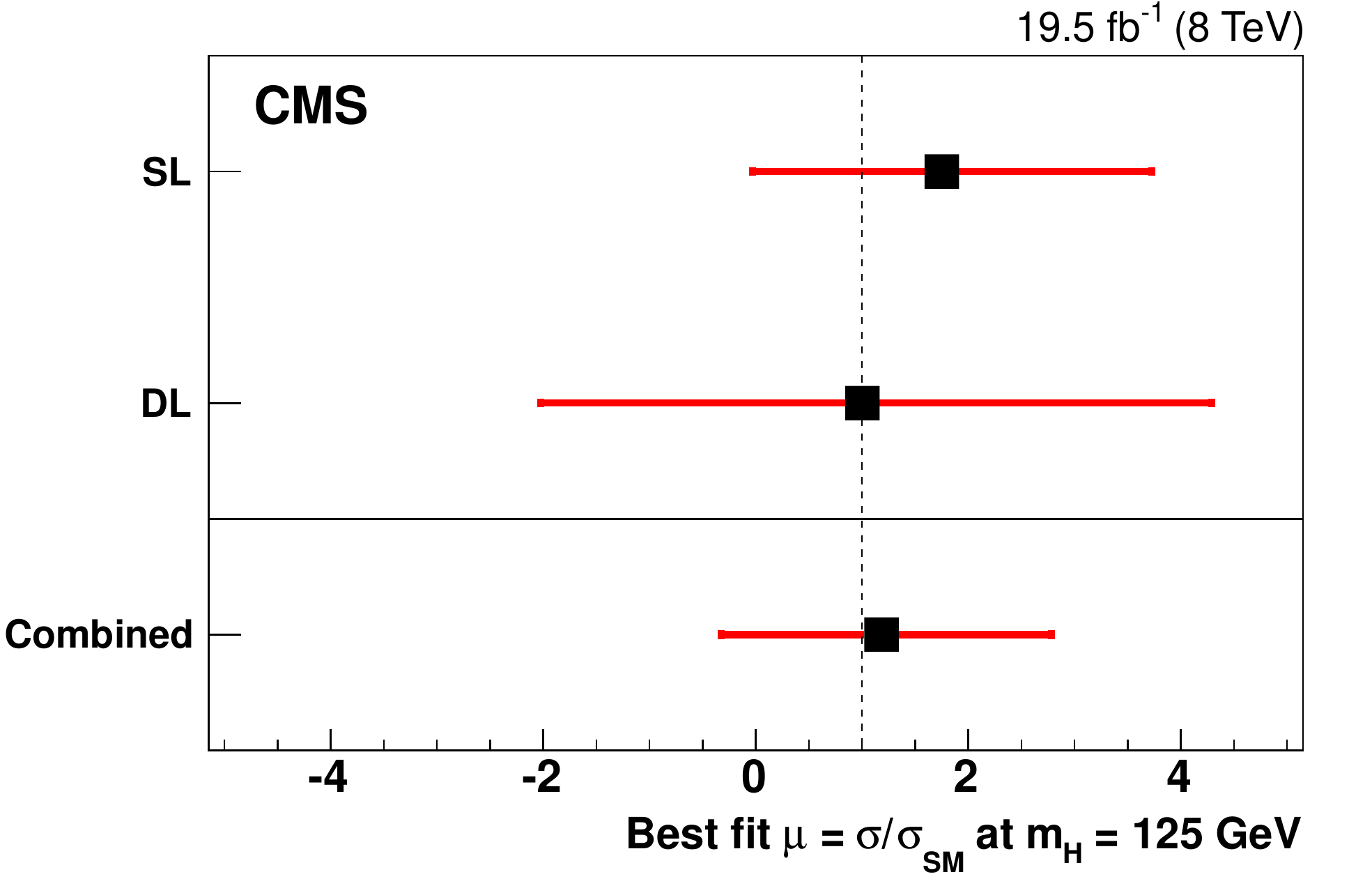}\\
  \includegraphics[width=0.475\textwidth]{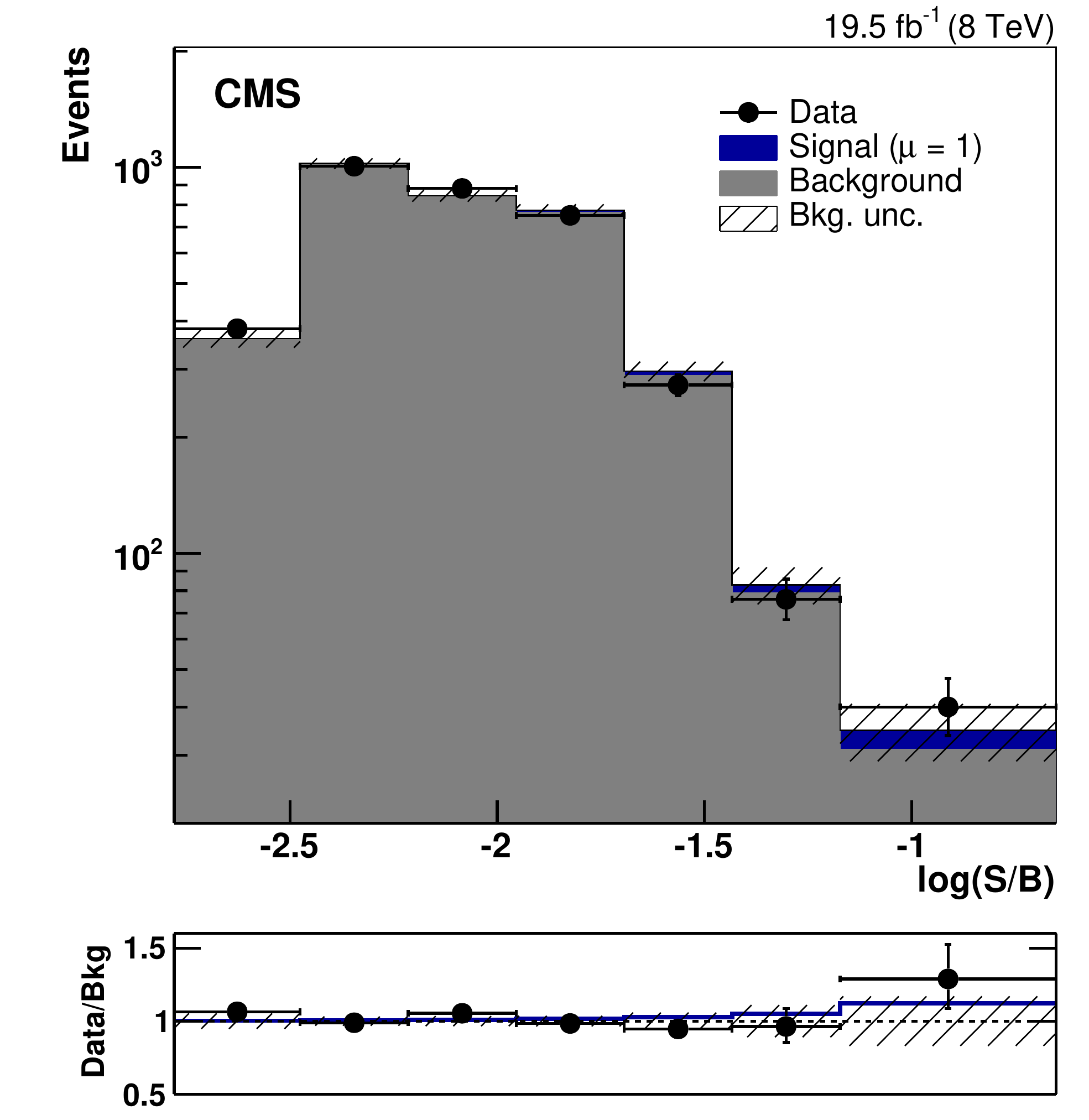}
\caption{
(\cmsTopLeft) Observed 95\% \CL UL on $\mu$ are compared to the median expected limits under the
background-only and the signal-plus-background hypotheses.
The former are shown together with their ${\pm}1\sigma$ and ${\pm}2\sigma$ \CL intervals.
Results are shown separately for the individual channels and for their combination.
(\cmsMiddle) Best-fit value of the signal strength modifier $\mu$ with its $\pm1\sigma$ \CL interval obtained from the individual channels and from their combination.
(bottom) Distribution of the decimal logarithm $\log(\mathrm{S}/\mathrm{B})$, where $\mathrm{S}$ ($\mathrm{B}$)
indicates the total signal (background) yield expected in the bins of the two-dimensional histograms,
as obtained from a combined fit with the constraint $\mu=1$.
}
\label{fig:limits}
\end{figure}

Figure~\ref{fig:limits} (\cmsTopLeft) shows the observed 95\% \CL UL on $\mu$,
compared to the signal-plus-background and to the background-only expectation.
Results are shown for the SL and DL channels alone, and for their combination.
The observed (background-only expected) exclusion limit is $\mu<4.2$ (3.3).
The best-fit value of $\mu$ obtained from the individual channels and from their combination is shown in Fig.~\ref{fig:limits} (\cmsMiddle).
A best-fit value $\hat{\mu}=1.2^{+1.6}_{-1.5}$ is measured from the combined fit.
Table~\ref{table:limits} summarises the results.

Overall, a consistent distribution of the nuisance parameters pulls is obtained from the combined fit.
In the signal-plus-background (background-only) fit, the nuisance parameters that account
for the 50\% normalisation uncertainty in the $\ttpBBbar$, $\ttpB$, and $\ttpCCbar$ backgrounds
are pulled by $+0.2$ ($+0.5$), $-0.4$ ($-0.3$), and $+0.8$ ($+0.8$), respectively, where the pull is defined
as the shift of the best-fit estimator from its nominal value in units of its a priori uncertainty.
The correlation between the $\ttpBBbar$ normalisation nuisance and the $\hat{\mu}$ estimator is found to be $\rho\approx-0.4$,
and is the largest entry in the correlation matrix.
From an a priori study (\ie before fitting the nuisance parameters with the likelihood function of the data),
the nuisance parameter corresponding to the 50\% normalisation uncertainty in the $\ttpBBbar$ background features the largest
impact on the median expected limit, which would be around 4\% smaller if that uncertainty were not taken into account.
Such a reduced impact on the expected limit implies that the sensitivity of the analysis is only mildly affected by
the lack of a stringent a priori constraint on the $\ttpBBbar$ background normalisation; this
is also consistent with the observation that the fit effectively constrains the $\ttpBBbar$ rate, narrowing
its normalisation uncertainty down to about 25\%.

\begin{table*}[htbp]
\topcaption{
The best-fit values of the signal strength modifier obtained from the SL and DL channels alone, and from their combination.
The observed 95\% \CL UL on $\mu$ are given in the third column, and are compared to the median expected
limits for both the signal-plus-background and for the background-only hypotheses.
For the latter, the $\pm1\sigma$ and $\pm2\sigma$ \CL intervals are also given.}
\centering
\ifthenelse{\boolean{cms@external}}{}{\resizebox{\textwidth}{!}}
{
\renewcommand{\arraystretch}{1.3}
\begin{tabular}{ *{7}{c} }
\multirow{2}{*}{Channel}   &  \multirow{2}{*}{Best-fit $\mu$}     & \multirow{2}{*}{Observed UL} &       Median exp. UL        & Median exp. UL & ${\pm}1\sigma$ \CL  & ${\pm}2\sigma$ \CL  \\
                                          &                                                      &                                                &      (signal injected)                  &       (background only)                               & interval     & interval      \\
\hline
 SL           & $+1.7^{+2.0}_{-1.8} $ & 5.5  & 5.0 & 4.2 & [2.9, 6.2]  & [2.1, 9.1] \\
 DL           & $+1.0^{+3.3}_{-3.0}$ & 7.7  & 7.8 & 6.9 & [4.7, 10.6] & [3.4, 15.8] \\
\hline
 Combined & $+1.2^{+1.6}_{-1.5}$ & 4.2 & 4.1 & 3.3 & [2.3, 4.9] & [1.7, 7.0] \\
\end{tabular}
}
\label{table:limits}
\end{table*}

For illustration, Fig.~\ref{fig:limits} (bottom) shows the distribution of the decimal logarithm $\log(\mathrm{S}/\mathrm{B})$,
where $\mathrm{S}/\mathrm{B}$ is the ratio between the signal and background yields in each bin of the two-dimensional histograms,
as obtained from a combined fit with the constraint $\mu=1$.
Agreement between the data and the SM expectation is observed over the whole range of this variable.

\section{Summary}\label{sec:end}

A search for Higgs boson production in association with a top-quark pair with $\HBB$ has been presented.
A total of 19.5\fbinv of $\Pp\Pp$ collision data collected by the CMS experiment at $\sqrt{s}=8\TeV$ has been analysed.
Events with one lepton and at least five jets or two opposite-sign leptons and at least four jets have been considered.
Jet \PQb-tagging information is exploited to suppress the \ttbar plus light-flavour background.
A probability density value under either the $\ttH$ or the $\ttpBBbar$ background hypothesis is calculated for each event
using an analytical matrix element method.
The ratio of probability densities under these two competing hypotheses allows a one-dimensional discriminant to be defined,
which is then used together with \PQb-tagging information in a likelihood analysis to set constraints on the signal strength modifier $\mu=\sigma/\sigma_{\mathrm{SM}}$.

No evidence of a signal is found. The expected upper limit at a 95\% \CL is $\mu<3.3$ under the background-only hypothesis.
The observed limit is $\mu<4.2$, corresponding to a best-fit value $\hat{\mu}=1.2^{+1.6}_{-1.5}$.
Within the present statistics, the analysis documented in this paper yields competitive results compared to those obtained on the same data set
and for the same final state by using non-analytical multivariate techniques~\cite{PAPERttHCMS}. However,
the matrix element method applied for a maximal separation between the signal and the dominant
$\ttpBBbar$ background allows for a better control of the systematic uncertainty due to this challenging background.
This method represents a promising strategy towards a precise determination of the top quark Yukawa coupling.
Once the statistical uncertainty will be reduced by the inclusion of the upcoming $13\TeV$ collision data,
systematic uncertainties will start to play a more important role.
By incorporating experimental and theoretical model parameters into an event likelihood,
the matrix element method offers a natural handle to minimise the impact of systematic
uncertainties on the extraction of the signal.

\begin{acknowledgments}
\hyphenation{Bundes-ministerium Forschungs-gemeinschaft Forschungs-zentren} We congratulate our colleagues in the CERN accelerator departments for the excellent performance of the LHC and thank the technical and administrative staffs at CERN and at other CMS institutes for their contributions to the success of the CMS effort. In addition, we gratefully acknowledge the computing centres and personnel of the Worldwide LHC Computing Grid for delivering so effectively the computing infrastructure essential to our analyses. Finally, we acknowledge the enduring support for the construction and operation of the LHC and the CMS detector provided by the following funding agencies: the Austrian Federal Ministry of Science, Research and Economy and the Austrian Science Fund; the Belgian Fonds de la Recherche Scientifique, and Fonds voor Wetenschappelijk Onderzoek; the Brazilian Funding Agencies (CNPq, CAPES, FAPERJ, and FAPESP); the Bulgarian Ministry of Education and Science; CERN; the Chinese Academy of Sciences, Ministry of Science and Technology, and National Natural Science Foundation of China; the Colombian Funding Agency (COLCIENCIAS); the Croatian Ministry of Science, Education and Sport, and the Croatian Science Foundation; the Research Promotion Foundation, Cyprus; the Ministry of Education and Research, Estonian Research Council via IUT23-4 and IUT23-6 and European Regional Development Fund, Estonia; the Academy of Finland, Finnish Ministry of Education and Culture, and Helsinki Institute of Physics; the Institut National de Physique Nucl\'eaire et de Physique des Particules~/~CNRS, and Commissariat \`a l'\'Energie Atomique et aux \'Energies Alternatives~/~CEA, France; the Bundesministerium f\"ur Bildung und Forschung, Deutsche Forschungsgemeinschaft, and Helmholtz-Gemeinschaft Deutscher Forschungszentren, Germany; the General Secretariat for Research and Technology, Greece; the National Scientific Research Foundation, and National Innovation Office, Hungary; the Department of Atomic Energy and the Department of Science and Technology, India; the Institute for Studies in Theoretical Physics and Mathematics, Iran; the Science Foundation, Ireland; the Istituto Nazionale di Fisica Nucleare, Italy; the Ministry of Science, ICT and Future Planning, and National Research Foundation (NRF), Republic of Korea; the Lithuanian Academy of Sciences; the Ministry of Education, and University of Malaya (Malaysia); the Mexican Funding Agencies (CINVESTAV, CONACYT, SEP, and UASLP-FAI); the Ministry of Business, Innovation and Employment, New Zealand; the Pakistan Atomic Energy Commission; the Ministry of Science and Higher Education and the National Science Centre, Poland; the Funda\c{c}\~ao para a Ci\^encia e a Tecnologia, Portugal; JINR, Dubna; the Ministry of Education and Science of the Russian Federation, the Federal Agency of Atomic Energy of the Russian Federation, Russian Academy of Sciences, and the Russian Foundation for Basic Research; the Ministry of Education, Science and Technological Development of Serbia; the Secretar\'{\i}a de Estado de Investigaci\'on, Desarrollo e Innovaci\'on and Programa Consolider-Ingenio 2010, Spain; the Swiss Funding Agencies (ETH Board, ETH Zurich, PSI, SNF, UniZH, Canton Zurich, and SER); the Ministry of Science and Technology, Taipei; the Thailand Center of Excellence in Physics, the Institute for the Promotion of Teaching Science and Technology of Thailand, Special Task Force for Activating Research and the National Science and Technology Development Agency of Thailand; the Scientific and Technical Research Council of Turkey, and Turkish Atomic Energy Authority; the National Academy of Sciences of Ukraine, and State Fund for Fundamental Researches, Ukraine; the Science and Technology Facilities Council, UK; the US Department of Energy, and the US National Science Foundation.

Individuals have received support from the Marie-Curie programme and the European Research Council and EPLANET (European Union); the Leventis Foundation; the A. P. Sloan Foundation; the Alexander von Humboldt Foundation; the Belgian Federal Science Policy Office; the Fonds pour la Formation \`a la Recherche dans l'Industrie et dans l'Agriculture (FRIA-Belgium); the Agentschap voor Innovatie door Wetenschap en Technologie (IWT-Belgium); the Ministry of Education, Youth and Sports (MEYS) of the Czech Republic; the Council of Science and Industrial Research, India; the HOMING PLUS programme of Foundation for Polish Science, cofinanced from European Union, Regional Development Fund; the Compagnia di San Paolo (Torino); the Consorzio per la Fisica (Trieste); MIUR project 20108T4XTM (Italy); the Thalis and Aristeia programmes cofinanced by EU-ESF and the Greek NSRF; and the National Priorities Research Program by Qatar National Research Fund.
\end{acknowledgments}
\bibliography{auto_generated}

\cleardoublepage \appendix\section{The CMS Collaboration \label{app:collab}}\begin{sloppypar}\hyphenpenalty=5000\widowpenalty=500\clubpenalty=5000\textbf{Yerevan Physics Institute,  Yerevan,  Armenia}\\*[0pt]
V.~Khachatryan, A.M.~Sirunyan, A.~Tumasyan
\vskip\cmsinstskip
\textbf{Institut f\"{u}r Hochenergiephysik der OeAW,  Wien,  Austria}\\*[0pt]
W.~Adam, T.~Bergauer, M.~Dragicevic, J.~Er\"{o}, M.~Friedl, R.~Fr\"{u}hwirth\cmsAuthorMark{1}, V.M.~Ghete, C.~Hartl, N.~H\"{o}rmann, J.~Hrubec, M.~Jeitler\cmsAuthorMark{1}, W.~Kiesenhofer, V.~Kn\"{u}nz, M.~Krammer\cmsAuthorMark{1}, I.~Kr\"{a}tschmer, D.~Liko, I.~Mikulec, D.~Rabady\cmsAuthorMark{2}, B.~Rahbaran, H.~Rohringer, R.~Sch\"{o}fbeck, J.~Strauss, W.~Treberer-Treberspurg, W.~Waltenberger, C.-E.~Wulz\cmsAuthorMark{1}
% ...
\vskip\cmsinstskip
% ...
1:~~Also at Vienna University of Technology, Vienna, Austria\\
2:~~Also at CERN, European Organization for Nuclear Research, Geneva, Switzerland\\
% ...

\end{sloppypar}
\end{document}